\documentclass[a4paper, twocolumn]{revtex4-1}

\usepackage[margin=1.8cm]{geometry}
\DeclareMathSizes{10}{9}{7}{5}

\usepackage{physics}
\usepackage{amssymb}
\usepackage{amsmath}
\usepackage{cancel}
\usepackage{empheq}

\usepackage{hyperref}
\hypersetup{colorlinks,citecolor=blue,urlcolor=blue,linkcolor=black,hypertexnames=true}

\newcommand{\Pf}{\mathcal{P}}
\newcommand{\E}{\mathcal{E}}
\newcommand{\Hf}{\mathcal{H}}

\begin{document}

\title{Optical momentum distributions in monochromatic, isotropic random vector fields}

\author{Titouan Gadeyne$^{1,2}$ and Mark R. Dennis$^{1,3,*}$}

\affiliation{$^1$ School of Physics and Astronomy, University of Birmingham, Birmingham, B15 2TT, UK}
\affiliation{$^2$ Département de Chimie, École Normale Supérieure, PSL University, 75005 Paris, France}
\affiliation{$^3$ EPSRC Centre for Doctoral Training in Topological Design, University of Birmingham, Birmingham, B15 2TT, UK}

\begin{abstract}
We investigate the decomposition of the electromagnetic Poynting momentum density in three-dimensional random monochromatic fields into orbital and spin parts, using analytical and numerical methods. 
In sharp contrast with the paraxial case, the orbital and spin momenta in isotropic random fields are found to be identically distributed in magnitude, increasing the discrepancy between the Poynting and orbital pictures of energy flow. 
Spatial correlation functions reveal differences in the generic organization of the optical momenta in complex natural light fields, with the orbital current typically forming broad channels of unidirectional flow, and the spin current manifesting larger vorticity and changing direction over subwavelength distances. 
These results are extended to random fields with pure helicity, in relation to the inclusion of electric-magnetic democracy in the definition of optical momenta.\\

\end{abstract}

\date{\today}

\maketitle

\email{$^*$\url{m.r.dennis@bham.ac.uk}}


\section{Introduction}

Conservation of electromagnetic (EM) energy is determined by the well-known theorem of Poynting \cite{poynting_1884_transfer}: in the absence of charges, the rate of change of EM energy density is equal to the divergence of the Poynting vector $\Pf = \E \cross \Hf$, the cross product of the electric and magnetic fields. 
By analogy with other continuity equations, it is customary to interpret $\Pf$ as the direction and magnitude of EM energy flow \cite{born_1980_principles, zangwill_2012_modern}. 
However, this choice often fails to produce an intuitive picture, even in seemingly elementary situations~: for instance, the Poynting vector for two crossed plane waves \cite{bekshaev_2015_transverse} or in a single evanescent surface wave \cite{fedorov_1955_theory, bliokh_2014_extraordinary} exhibits a counterintuitive component \emph{perpendicular} to the direction of propagation. 
Similarly, the (time-averaged) radiation pressure exerted by an optical field on a subwavelength probe particle is generally \emph{not} proportional to the Poynting vector \cite{CanaguierDurandForce2013, sukhov_2017_non}. 

Divided by $c^2$, the Poynting vector also defines the linear momentum density of the EM field. 
It is now well understood that in monochromatic fields, the time-averaged linear momentum $\vb{P}$ is the sum of \emph{orbital} $\vb{P}_O$ and \emph{spin} $\vb{P}_S$ parts, respectively generating the orbital and spin angular momenta \cite{bekshaev_2007_transverse, berry_2009_optical, bekshaev_2011_internal}. 
In \cite{berry_2009_optical}, these vector fields were dubbed \emph{optical currents}. 
This Poynting vector splitting has deep foundations, as the orbital momentum in fact corresponds to the canonical momentum density derived from application of Noether's theorem to translational invariance in the relativistic field theory formulation of electromagnetism \cite{jackson_1999_classical,bliokh_2013_dual}.                                      
The orbital momentum correctly accounts for the radiation pressure on dipole particles, and can provide a more intuitive picture of energy flow than the Poynting vector in the situations mentioned above. 
In the field theory framework, the spin momentum corresponds to a term introduced by Belinfante \cite{belinfante_1940_current} to guarantee symmetry and gauge-invariance to the EM stress-energy tensor \cite{jackson_1999_classical}, which does not contribute to the total linear momentum of the field when integrated over space. 
As such, the Belinfante \emph{spin momentum} is often described as a ``virtual'' quantity introduced for theoretical reasons. 
Nevertheless, this spin momentum has recently been evidenced experimentally, by measuring the extraordinary optical force it induced on a nano-cantilever \cite{antognozzi_2016_direct}. 
Importantly, the couplings to the orbital and spin parts of the Poynting vector differ by orders of magnitude, highlighting their distinct physical nature. 

Recent experimental and theoretical studies have thus demonstrated striking differences between the Poynting, orbital, and spin momenta, and continue to redefine our views of EM energy flow and optical forces \cite{bekshaev_2011_internal, sukhov_2017_non, nietovesperinas_2022_complex}. 
Still, they have so far been limited to rather elementary, highly symmetric fields, with geometries optimized to best showcase the differences between the three optical currents. 
In particular, the focus on light \emph{beams} ensures the orbital momentum, related to the propagation $k$-vector of the field, is typically large, whereas the spin momentum, related to the rate of change of polarisation with position, is smaller.

In this work, we explore \emph{generic} features of these optical currents, to build insight into their organization in \emph{natural, isotropic light fields}~: what are their properties when many independent waves interfere with no particular symmetries~? 
To this end, we investigate their behaviour in completely coherent, monochromatic, isotropic random EM vector fields, a convenient statistical model of 3D EM fields specified by only one physical parameter, the wavelength $\lambda$. 
Strikingly, we will see that in this model, the magnitudes of the spin and orbital currents have the same probability distribution, but that the two vector fields have different spatial correlations~: the apparent weakness of the spin current in this setting is due to its failure to organise coherent correlated vector structures over large distances in space, unlike the orbital current (and Poynting vector itself). 
We demonstrate these facts using analytical statistics and numerical simulations for the vector random wave model.

\section{Theoretical framework}

\subsection{Poynting, orbital and spin momenta}

We work in units where $\varepsilon_0=\mu_0=c=1$. 
In a monochromatic field with frequency $\omega = ck = c(2\pi/\lambda)$, the electric and magnetic fields are represented by complex-valued vector fields $\vb{E}$ and $\vb{H}$, where the physical fields are $\mathcal{E} = \Re \{\vb{E} e^{-i\omega t} \}$ and $\mathcal{H} = \Re \{\vb{H} e^{-i\omega t} \}$. 
The temporal cycle-averaged Poynting momentum can be written
\begin{align}
    \vb{P} = \frac{1}{2} \Re{\vb{E}^* \cross \vb{H}}.
\end{align}
Using Maxwell's equations 
and the vector identity $\vb{A} \cross (\curl{\vb{B}}) = \vb{A} \vdot (\grad) \vb{B} - (\vb{A} \vdot \grad) \vb{B}$ (where we use the customary notation $[\vb{A} \vdot (\grad) \vb{B} ]_i = \sum_j A_j \partial_i B_j$), the Poynting momentum can be split into a sum of orbital and spin momenta \cite{berry_2009_optical},
\begin{align}
\begin{split}
    \vb{P} 
    & =
    \frac{1}{2 \omega} \Im{\vb{E}^* \vdot (\grad) \vb{E}}
    -
    \frac{1}{2 \omega} \Im{(\vb{E}^* \vdot \grad) \vb{E}}
    \\
    & =
    \vb{P}^{\vb{E}}_O + \vb{P}^{\vb{E}}_S.
\end{split}
\label{eq:biased_splitting}
\end{align}
This splitting is not unique~:~expressing the Poynting momentum using the electric (resp.~magnetic) field only, we obtain the \emph{electric-biased} (resp.~\emph{magnetic-biased}) momenta $\vb{P}^{\vb{E}}_{O,S}$ (resp.~$\vb{P}^{\vb{H}}_{O,S}$) as in \eqref{eq:biased_splitting}. 
But another option, retaining the electric-magnetic symmetry of Maxwell's equations for free fields, is to take the mean of the two representations,
\begin{align}
\begin{split}
    \vb{P} 
    & =
     \frac{1}{4 \omega} \Im{
    \vb{E}^* \vdot (\grad) \vb{E}
    +
    \vb{H}^* \vdot (\grad) \vb{H}
    }
    \\
    & -
    \frac{1}{4 \omega} \Im{
    (\vb{E}^* \vdot \grad) \vb{E}
    +
    (\vb{H}^* \vdot \grad) \vb{H}
    }
    \\
    & =
     \vb{P}^{\vb{EH}}_O + \vb{P}^{\vb{EH}}_S,
\end{split}
\label{eq:democratic_splitting}
\end{align}
producing the so-called \emph{democratic} (or \emph{dual}) momenta $\vb{P}^{\vb{EH}}_O,\vb{P}^{\vb{EH}}_S$ \cite{berry_2009_optical}. 
In general non-paraxial fields, these are all distinct quantities, and in monochromatic fields, their definition is unambiguous (otherwise the splitting is gauge-dependent). 
An interesting situation arises in fields with \emph{pure helicity}, consisting only of circularly-polarized plane wave components of same handedness~:~such fields satisfy $\vb{E} = \pm i \vb{H}$, such that all biased and democratic quantities become identical.

The dual formulation of electromagnetism that treats electric and magnetic fields equally has many attractive features when working with free fields, in the absence of matter \cite{bliokh_2013_dual, CameronElectricmagnetic2012} --- for instance, democratic momenta naturally split into two independent parts associated with components of opposite helicity \cite{aiello_2015_note}. 
However, experimental measurements require material probes, which typically do \emph{not} respond identically to electric and magnetic fields~:~a common example is the radiation pressure on a subwavelength particle responding in the electric dipole approximation, which is proportional to the \emph{electric-biased} orbital momentum $\vb{P}^{\vb{E}}_O$ only.
We therefore choose to center our discussion on electric-biased quantities, and devote section \ref{subsec:democratic} to observations on democratic momenta and pure helicity fields for which the distinction vanishes.

For reference, we briefly recall the typical magnitudes of the three momenta in a paraxial beam \cite{bekshaev_2007_transverse, bekshaev_2011_internal} propagating along $z$, for which the field is approximately $\vb{E} \approx e^{i k z} (E_x, E_y, 0)$ with transverse amplitude and polarization profiles $E_x(x,y)$ and $E_y(x,y)$ varying over a lengthscale $W \gg \lambda$, on the order of the beam waist. 
We find that $\vb{P}$ and $\vb{P}^{\vb{E}}_O$ are mostly longitudinal, that $\vb{P}^{\vb{E}}_S$ is purely transverse, and the following orders of magnitude
\begin{align*}
    \abs{\vb{P}} & \sim E^2,
    \\
    \abs{\vb{P}^{\vb{E}}_O} & \sim \frac{1}{\omega}\Im{ E^* \partial_z E} \sim \frac{k}{\omega}E^2 \sim E^2,
    \\
    \abs{\vb{P}^{\vb{E}}_S} & \sim \frac{1}{\omega}\Im{ E^* \partial_x E} \sim \frac{\lambda}{W}E^2 \ll E^2.
\end{align*}
We conclude that in a regular optical beam, orbital and Poynting momenta are closely aligned, the spin momentum being small in comparison. 

\subsection{Gaussian random optical fields}

We model generic, natural EM light fields as superpositions of $N \rightarrow +\infty$ plane waves, with randomly sampled propagation directions taken uniformly over the sphere of directions, random elliptical polarizations, and random overall phases. 
This construction aims to portray an unprepared field, akin to ambient light in a room, or thermal black-body radiation, composed of many waves emitted from independent points or having scattered off various surfaces, producing a field with no particular symmetries or preferred directions, but statistically homogeneous and isotropic. 
This approach builds on the long history of the study of speckle patterns \cite{goodman_1985_statistical, goodman_2007_speckle, freund_1998_1001} and statistical properties of spatial structures in random light fields \cite{berry_2000_phase, berry_2001_polarization, dennis_2002_polarizationc, dennis_2007_nodal, berry_2019_geometry}, which revealed salient features underlying the organization of all EM fields. 
Physically and geometrically, these random vector fields are very different from directed, paraxial optical beams.
\begin{figure*}[ht!]
    \centering
    \includegraphics[width=\textwidth, trim={0, 1cm, 0, 0}]{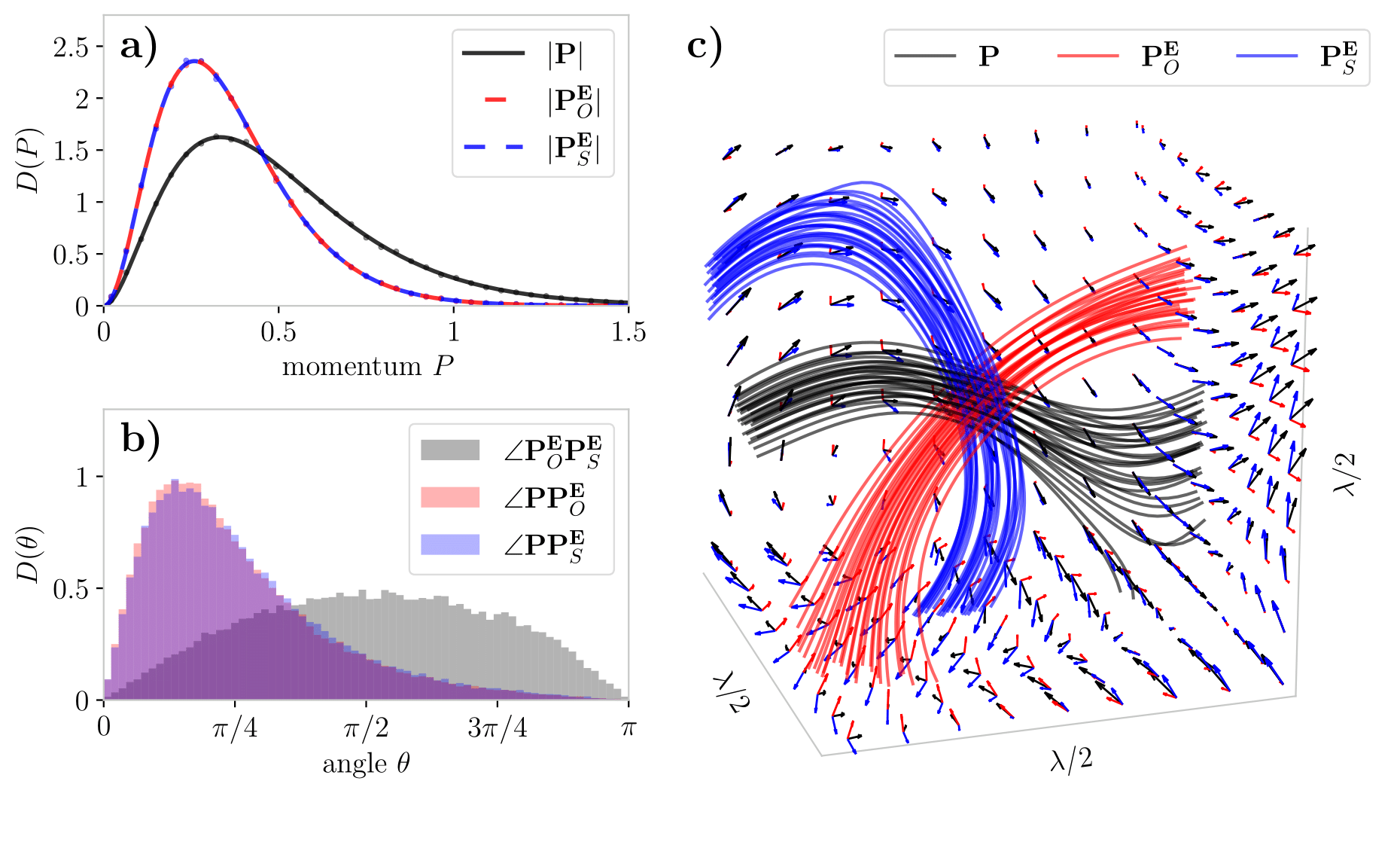}
    \caption{\textbf{Magnitudes and relative orientations of the Poynting, orbital and spin momenta.} \\
    (a) Analytical (lines) and numerical (dots) probability distributions for the magnitudes of the Poynting, orbital and spin momenta. 
    (b) Distribution of the angles between momenta, obtained numerically. 
    (c) Illustration of one realisation of the random field, showing the three vector fields on the faces of a cubic region of side $\lambda/2$, and a set of streamlines seeded near the center of the cube. 
    Numerical data in this figure was obtained from $10^5$ realizations of the random field, each containing $N=10^3$ plane waves. }
    \label{fig:1}
\end{figure*}
The complex electric and magnetic fields can be parameterized as follows,
\begin{subequations}\label{eq:plane_waves}
\begin{equation}
\begin{split}
    \vb{E} = 
    \sqrt{\frac{2}{N}} \sum_{n=1}^N
    e^{i \vb{k}_n \vdot \vb{r} + i \psi_n}
    \Big[ 
    e^{i\alpha_n/2} \cos\frac{\beta_n}{2} \vb{e}_{+}(\vb{k}_n)&
    \\
    + e^{-i \alpha_n/2} \sin\frac{\beta_n}{2} \vb{e}_{-}(\vb{k}_n)&
    \Big],
\end{split}
\end{equation}
\begin{equation}
\begin{split}
    \vb{H} = 
    \sqrt{\frac{2}{N}} \sum_{n=1}^N
    e^{i \vb{k}_n \vdot \vb{r} + i \psi_n}
    \frac{1}{i}
    \Big[  
    e^{i\alpha_n/2} \cos\frac{\beta_n}{2} \vb{e}_{+}(\vb{k}_n)&
    \\
    - e^{-i \alpha_n/2} \sin\frac{\beta_n}{2} \vb{e}_{-}(\vb{k}_n)&
    \Big],
\end{split}
\end{equation}
\end{subequations}
where the sum runs over the $N\gg 1$ plane waves, with wavevectors $\vb{k}_n$ sampled uniformly on the sphere of directions with spherical angles ($\theta_n$, $\phi_n$) and identical magnitudes $k$, and polarizations sampled uniformly on the Poincaré sphere with angles ($\beta_n$, $\alpha_n$), and uniformly sampled global phases $\psi_n$. 
$\vb{e}_{\pm}(\vb{k}) = [\vb{e}_1 \pm i \vb{e}_2]/\sqrt{2}$ are helicity basis vectors, with \{$\vb{e}_1, \vb{e}_2$\} a basis of two real orthogonal unit vectors transverse to $\vb{k}$ (see the SI for explicit expressions and alternative parameterizations). 
We introduce the following notation for the real and imaginary parts of the fields \cite{born_1980_principles, berry_2019_geometry}
\begin{align*}
    \vb{E} = \vb{p}^{\vb{E}} + i \vb{q}^{\vb{E}}, 
    \;\;\;\;\;
    \vb{H} = \vb{p}^{\vb{H}} + i \vb{q}^{\vb{H}},
\end{align*}
since statistics are convenient with real quantities only. Ensemble-averaging over many random fields is denoted by brackets and amounts to integrating over the five random angles
\begin{equation}
\begin{split}
    \expval{\bullet}
    = 
    \prod_{n=1}^N 
    \Bigg[
    & \frac{1}{32 \pi^3}
    \int_0^\pi \sin\theta_n \dd \theta_n
    \int_0^{2\pi} \dd \phi_n
    \\
    & \;
    \times
    \int_0^\pi \sin\beta_n \dd \beta_n
    \int_0^{2\pi} \dd \alpha_n
    \int_0^{2\pi} \dd \psi_n
    \Bigg]
    \bullet.
\end{split}
\label{eq:averaging}
\end{equation}
From the definitions above, it can be seen that any component of the real or imaginary part of a field is a sum of $N$ real-valued, identically distributed random variables. 
The central limit theorem ensures that in the limit $N \rightarrow +\infty$, each component is a real random variable obeying Gaussian statistics \cite{goodman_1985_statistical, goodman_2007_speckle}. 
The same reasoning holds for all derivatives of the components. 
In our case these variables are all centered, hence we only require their variances and correlations to fully describe the statistics. 
They are obtained by direct integration using \eqref{eq:averaging}, and are tabulated in the SI. 
With these provided, an ensemble average rewrites as an integral over a set of $M$ Gaussian random variables $\vb{u} = (p_x^{\vb{E}}, p_y^{\vb{E}} \ldots)$
\begin{align}
    \expval{\bullet}
    =
    \sqrt{\frac{\det{\vb{\Sigma^{-1}}}}{(2\pi)^M}}
    \int \ldots \int \dd^M \vb{u} \exp{-\frac{\vb{u}^\intercal \vb{\Sigma^{-1}} \vb{u}}{2} }
    \bullet,
\label{eq:Gauss_averaging}
\end{align}
where $\vb{\Sigma}$ is the covariance matrix, with $\Sigma_{ij} = \expval{u_i u_j}$. 
Useful formulae and strategies for computing averages are further described in the SI, and can be found in references \cite{berry_2000_phase, berry_2001_polarization, dennis_2003_correlations, dennis_2007_nodal, berry_2019_geometry}. 

\subsection{Spatial correlation functions}

To investigate local order in the spatial organization of the optical currents, we will average products of vector components at two different positions in space. 
The statistical, directional correlators in random Gaussian vector fields, here representing EM waves, are analogous to those used in the theory of isotropic turbulence in fluids \cite{batchelor_1953_theory}.
For a homogeneous random vector field $\vb{v}$ to be isotropic requires the two-point correlation tensor to have the form
\begin{align*}
    \expval{v_i(\vb{0}) v_j(\vb{r})}
    =
    [f(r)-g(r)]\frac{r_i r_j}{r^2} + g(r) \delta_{ij},
\end{align*}
where $f$ and $g$ are scalar functions depending only on the magnitude $r = |\vb{r}|$ of the separation vector. They respectively describe \emph{longitudinal} and \emph{lateral} autocorrelations of a given vector component
\begin{align*}
    f(r) = \expval{v_i(\vb{0}) v_i( r \vb{e}_i)},
    \;\;\;\;\;
    g(r) & = \expval{v_i(\vb{0}) v_i( r \vb{e}_j)} \; (i \neq j).
\end{align*}
where the separation vector $r\vb{e_i}$ is taken along some chosen direction $i=x,y,z$.
If, in addition, the field is solenoidal $(\div{\vb{v}} = 0)$, $f$ and $g$ are related, such that the full correlation tensor can be determined from, for example, the longitudinal correlation function $f$ only,
\begin{align}
    \expval{v_i(\vb{0}) v_j(\vb{r})}
    =
    - \frac{r f'(r)}{2} \frac{r_i r_j}{r^2} + \delta_{ij} \left[ f(r) + \frac{r f'(r)}{2} \right].
\label{eq:isotropic_solenoidal_tensor}
\end{align}

Since there are no charges in the model field, cycle-averaging Poynting's theorem yields $\div{\vb{P}} = 0$. 
As the spin momentum itself is the curl of a vector field \cite{berry_2009_optical}, it is divergenceless $\div{\vb{P}_S}=0$, and consequently we also have $\div{\vb{P}_O} = 0$. 
Hence all momenta are isotropic homogeneous solenoidal random fields, to which the above results apply. 
They also apply to the complex electric and magnetic fields themselves. In our calculations, we will be able to express all correlation functions using the longitudinal and lateral autocorrelation functions of the electric field \cite{bourret_1960_coherence}, that we respectively denote $L$ and $T$
\begin{align*}
    L(r) & = \expval{p^{\vb{E}}_x(\vb{0})p^{\vb{E}}_x(r \vb{e}_x)}
    =
    \frac{\sin(R) - R \cos(R)}{R^3}
    \\
    T(r) & = \expval{p^{\vb{E}}_x(\vb{0})p^{\vb{E}}_x(r \vb{e}_y)}
    =
    \frac{R \cos(R) - (1 - R^2) \sin(R)}{2 R^3},
\end{align*}
where $R = kr$. 
Further useful strategies and elementary correlation functions are provided in the SI.

\section{Results and discussion}

All analytical derivations can be found in great detail in the SI. 
We mostly state final results here, except when intermediate steps are useful for understanding how a result comes about.

\subsection{Magnitudes and relative directions of the optical momenta}

We begin by deriving the fundamental statistical distributions for the magnitudes of the Poynting, orbital and spin momenta. 
In terms of real and imaginary field components, the Poynting momentum writes
\begin{align*}
    \vb{P} =
    \frac{1}{2} \left[ \vb{p^{\vb{E}}} \cross \vb{p}^{\vb{H}} + \vb{q}^{\vb{E}} \cross \vb{q}^{\vb{H}}  \right].
\end{align*}
Each component of $\vb{P}$ is a sum of products of two Gaussian random variables. 
As detailed in the SI, isotropy allows us to retrieve the magnitude distribution $D(P)$ from that of the $x$-component $D_x(P_x)$ only \cite{berry_2019_geometry}.  
We briefly outline this first derivation, to see the main steps involved:
\begin{align*}
    D_x(P_x) & = \expval{\delta \Big(P_x - \sum_{j,k} \frac{\epsilon_{xjk}}{2} \left[ p^{\vb{E}}_j p^{\vb{H}}_k + q^{\vb{E}}_j q^{\vb{H}}_k  \right] \Big)}
    \\
    & =
    \int \frac{\dd s}{2\pi} e^{-isP_x} \expval{\exp{i s \frac{1}{2} p_y^{\vb{E}} p_z^{\vb{H}}}}^4
    \\
    & =
    \int \frac{\dd s}{2\pi} e^{-isP_x} 
    \left[ 
    \frac{1}{1 + s^2 \sigma_x^4 / 4}
    \right]^2
    \\
    & =
    \frac{1 + 2|P_x|/\sigma_x^2}{2\sigma_x^2}
    \exp{-\frac{2 |P_x|}{\sigma_x^2}}, 
\end{align*}
where $\sigma_x^2 = \expval{(p^{\vb{E}}_x)^2} = 1/3$ (see tabulated variances in the SI). 
The second step involves factorization of the average using the statistical independence of field components, the third step uses \eqref{eq:Gauss_averaging}, and the last step is an integration in the complex plane. 
The distribution for the magnitude of the Poynting momentum is then
\begin{align*}
\begin{split}
    D(P) & = -2 P \pdv{D_x(P_x)}{P_x}\Big|_{P_x=P}
    =
    108 P^2 \exp{-6P}. 
\end{split}
\end{align*}
The electric-biased orbital and spin momenta read
\begin{align*}
    \vb{P}^{\vb{E}}_O & = 
    \frac{1}{2\omega} \vb{p}^{\vb{E}} \vdot ( \grad )\vb{q}^{\vb{E}} 
    -
    \frac{1}{2\omega} 
    \vb{q}^{\vb{E}} \vdot ( \grad )\vb{p}^{\vb{E}} 
    \\
    \vb{P}^{\vb{E}}_S & =
    -\frac{1}{2\omega} (\vb{p}^{\vb{E}} \vdot \grad) \vb{q}^{\vb{E}} 
    +
    \frac{1}{2\omega} 
    (\vb{q}^{\vb{E}} \vdot \grad) \vb{p}^{\vb{E}}. 
\end{align*}
Again, each component is a sum of products of two Gaussian random variables (one field component and one space derivative), and we only have to find the distribution for the $x$-component. 
For the orbital momentum this is
\begin{align*}
    & D_x(P^{\vb{E}}_{O,x}) = \expval{\delta \Big(P^{\vb{E}}_{O,x} - \sum_{j} \frac{1}{2}\left[ p^{\vb{E}}_j \partial_x q^{\vb{E}}_j - q^{\vb{E}}_j \partial_x p^{\vb{E}}_j  \right] \Big)} \\
    & =
    \int \frac{\dd s}{2\pi} e^{-isP^{\vb{E}}_{O,x}} \expval{\exp{ \frac{i s}{2} p^{\vb{E}}_x \partial_x q^{\vb{E}}_x}}^2
    \expval{\exp{ \frac{i s}{2} p^{\vb{E}}_y \partial_x q^{\vb{E}}_y}}^4
    \\
    & = \ldots
\end{align*}
and the distribution for the magnitude is
\begin{align*}
    D(P^{\vb{E}}_{O}) 
    & =
    180 P^{\vb{E}}_{O}
    \left[
    e^{-6\sqrt{5} P^{\vb{E}}_{O}}
    - \left[ 1
    -\frac{3\sqrt{10}}{2}
    P^{\vb{E}}_{O}
    \right]
    e^{-3 \sqrt{10} P^{\vb{E}}_{O}}
    \right].
\end{align*}
\begin{figure*}[t!]
    \centering
    \includegraphics[width=\textwidth, trim={0, 0.4cm, 0, 0}]{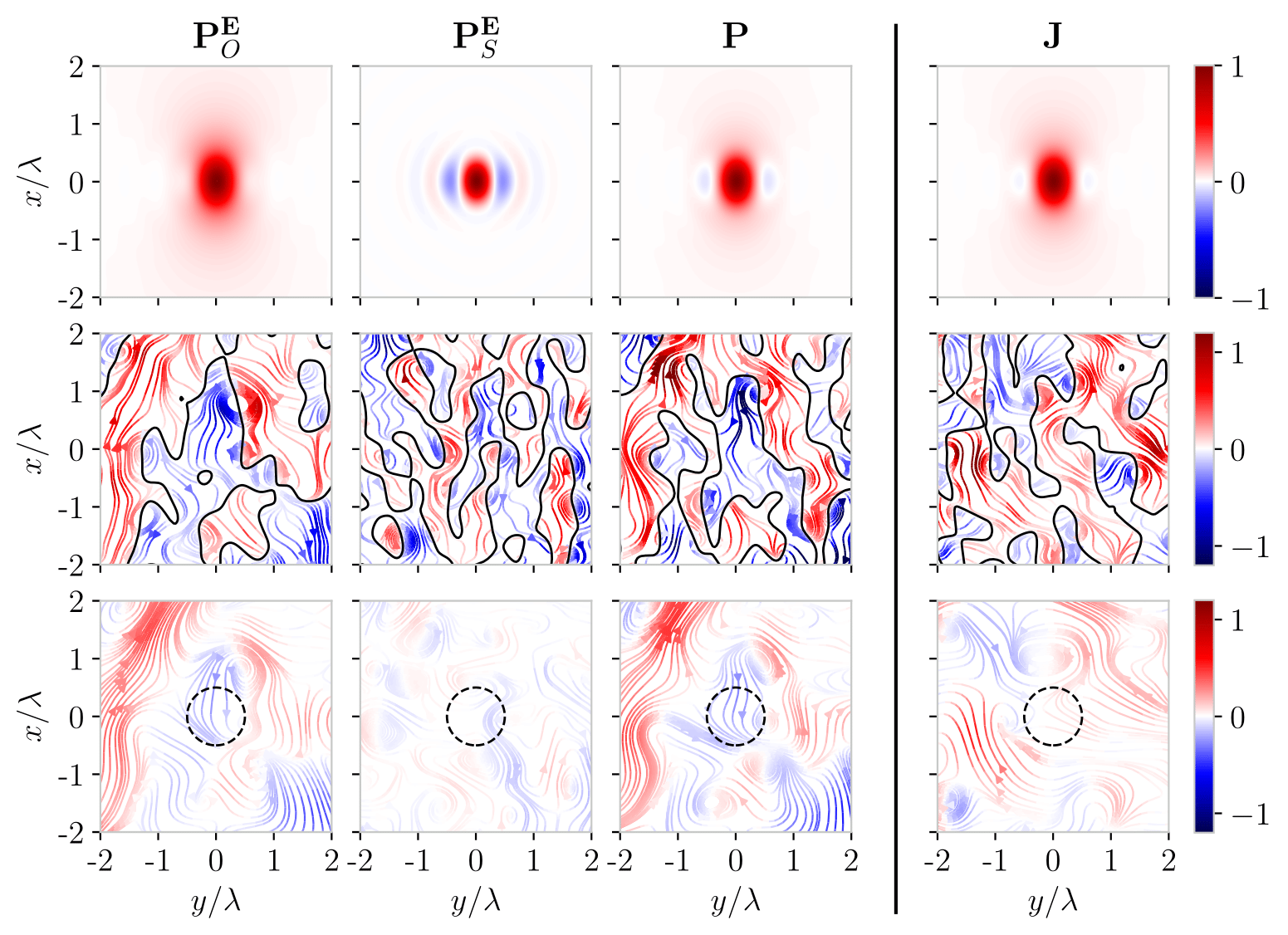}
    \caption{\textbf{Local structure of the optical currents.} \\
    First three columns~: optical currents of the vector EM field $\vb{V} = \vb{P}^{\vb{E}}_O,~\vb{P}^{\vb{E}}_S,~\vb{P}$. 
    Last column~: current in the complex scalar field $\vb{V} = \vb{J}$.
    First row~: normalized analytical spatial autocorrelation functions $\expval{V_x(\vb{0}) V_x(\vb{r})}/\expval{(V_x)^2}$ of the $x$-component of each momentum, for separation vectors $\vb{r}$ in the $x-y$ plane. 
    Second row~: in-plane streamlines of each momentum in a slice through one realization of the random vector field (first three columns) and one realization of the random scalar field (last column), each containing $N=10^3$ plane waves. 
    Streamlines are colored according to the value of the $x$-component of the vector field, and zero-crossings are shown in black to better distinguish regions having a flow oppositely directed along $x$. 
    Third row~: in-plane streamlines in the same slices as in the second row, after local averaging of the vector fields over a spherical volume of diameter $\lambda$ (dashed circle). 
    All plots in a given row share the same colorbar.}
    \label{fig:2}
\end{figure*}
Surprisingly, we find that in repeating the calculation for the spin momentum, the result is the same. 
Indeed, the first steps of the derivation read
\begin{align*}
    & D_x(P^{\vb{E}}_{S,x}) = \expval{\delta \Big(P^{\vb{E}}_{S,x} - \sum_{j} \frac{1}{2}\left[ p^{\vb{E}}_j \partial_j q^{\vb{E}}_x - q^{\vb{E}}_j \partial_j p^{\vb{E}}_x  \right] \Big)} \\
    & =
    \int \frac{\dd s}{2\pi} e^{-isP^{\vb{E}}_{S,x}} \expval{\exp{ \frac{i s}{2} p^{\vb{E}}_x \partial_x q^{\vb{E}}_x}}^2
    \expval{\exp{\frac{i s}{2} p^{\vb{E}}_y \partial_y q^{\vb{E}}_x}}^4
\end{align*}
and since $\partial_x q^{\vb{E}}_y$ and $\partial_y q^{\vb{E}}_x$ are both uncorrelated to $p^{\vb{E}}_y$ and have the same variance (see tables in the SI), the rest of the calculation is strictly identical to that for the orbital momentum, and we conclude that the orbital and spin momenta obey the exact same magnitude distribution. 
All these distributions are wavelength-independent, and only scale with the overall intensity in the field. 

They are plotted and checked against numerical estimates in \autoref{fig:1}.a). 
It is interesting to observe that the spin momentum, usually negligibly small in paraxial beams, becomes here equivalent in magnitude to the orbital momentum, responsible for the actual energy flow. 
The intuitive reason for the different order of magnitude is that in the fully non-paraxial case, there are waves propagating in all directions, such that all space derivatives result in a factor $\sim i k$, whereas transverse gradients are only of order $\sim 1/W$ in the paraxial case. 
Moving away from paraxiality, part of the linear momentum converts from an orbital to a spin nature, in a manner strictly similar to how the \emph{angular} momentum does \cite{bliokh_2010_angular}.

To complete the picture of the three momenta at a given point in space, we present in \autoref{fig:1}.b) the distributions for the angles between each pair of momenta. 
They were obtained numerically, as attempting to compute analytical joint distributions of two momenta hardly leads to tractable expressions. 
We observe that the angle between $\vb{P}^{\vb{E}}_O$ and $\vb{P}^{\vb{E}}_S$ has a broad distribution roughly centered on $\pi/2$ (with a slight skew towards larger angles), indicating that they tend to point in perpendicular directions. 
Since they have comparable magnitudes, the resulting Poynting momentum is generically not closely aligned with either of them. 
This implies that the streamlines for the three optical currents tend to diverge away from one another. 

In \autoref{fig:1}.c), we illustrate one realisation of the random field, in a cubic region of side $\lambda/2$. 
We plot the three momenta on the sides of the box, and a set of streamlines seeded near the center of the cube. 
The three vector fields are indeed observed to generically point in different directions, and the streamlines to follow seemingly unrelated paths in space, crossing with angles in agreement with the distributions of \autoref{fig:1}.b). 
This reinforces the claim that the Poynting and orbital currents generally provide contrasting pictures of EM energy flow, both in terms of magnitude and direction. These observations could prove important for simulating optical forces in complex nanophotonics systems.

\subsection{Short-range organization of the currents}

\subsubsection{Spatial correlation tensors}

Going beyond their identical magnitude distribution, we find that the orbital and spin momentum vector fields are actually arranged very differently in space. 
To explore this, we compute two-point spatial correlation tensors for all pairs of components of a given momentum. 
Each tensor will be of the form in \eqref{eq:isotropic_solenoidal_tensor}, given entirely by the longitudinal autocorrelation function $f(r)$. 
For the Poynting momentum, this function writes 
\begin{align*}
    & f_P(r) = \expval{P_x(\vb{0}) P_x(r\vb{e_x})}
    \\
    & = 
    \sum_{j,k,l,m}
    \expval{ \frac{1}{4} \epsilon_{xjk} \epsilon_{xlm} 
    [p^{\vb{E}}_j p^{\vb{H}}_k + q^{\vb{E}}_j q^{\vb{H}}_k](\vb{0})
    [p^{\vb{E}}_l p^{\vb{H}}_m + q^{\vb{E}}_l q^{\vb{H}}_m](r\vb{e}_x)
    }.
\end{align*}
To evaluate these averages we make use of Isserlis' theorem for moments of Gaussian variables \cite{goodman_1985_statistical}.
$f_P(r)$ is obtained as
\begin{align*}
    f_P(r) 
    & = 
    T^2(r) + \frac{(kr)^2}{4} L^2(r),
\end{align*}
and the correlation tensor is
\begin{align*}
\begin{split}
    & \expval{P_i(\vb{0}) P_j(\vb{r})} 
    =
    \frac{r_i r_j}{r^2} \Big[
    2 R \left(R^2-3\right) \sin (2 R) \\
    & +\left(6 R^2-3\right) \cos (2 R) +2 R^4 +3
    \Big] \Big/ 8 R^6
    \\
    &+ 
    \delta_{ij} \Big[
    R \left(2-R^2\right) \sin (2 R)
    +\left(1-2 R^2\right) \cos (2R)-1
    \Big] \Big/ 4 R^6.
\end{split}
\end{align*}
The strategy is similar for the orbital and spin momenta. 
We find
\begin{align*}
    \omega^2 f_O(r) 
     & = 
    \frac{1}{2}
    \left[
    {L'}^2(r)
    -
    L(r)L''(r)
    \right]
    +
    \left[
    {T'}^2(r)
    -
    T(r)T''(r)
    \right],
\end{align*}
giving the correlation tensor
\begin{align*}
\begin{split}
    & \expval{P^{\vb{E}}_{O,i}(\vb{0}) P^{\vb{E}}_{O,j}(\vb{r})} =
    \frac{r_i r_j}{r^2}
    \Big[
     \frac{1}{2} \left(R^4-24 R^2+72\right) R \sin (2 R) 
    \\
    & +3 \left(R^4-10 R^2+6\right) \cos (2 R) + R^6-3 R^4-6 R^2 - 18
    \Big] 
    \Big/ 4 R^8
    \\
    & 
    + \delta_{ij}
    \Big[ 
    -\left(R^4-20 R^2+54\right) R \sin (2 R)
    \\
    & \;\;\;\;\;
    +\left(-5 R^4+46 R^2-27\right) \cos (2 R) + 3 R^4+8 R^2+27
    \Big]
    \Big/ 8 R^8.
\end{split}
\end{align*}
And for the spin momentum,
\begin{align*}
    \omega^2 f_S(r) 
    & = 
    \frac{3}{4}
    {L'}^2(r)
    - \frac{L(r) L''(r)}{2}
    -
    2 L'(r)
    \frac{T(r)}{r}
\end{align*}
with the correlation tensor
\begin{align*}
\begin{split}
    & \expval{P^{\vb{E}}_{S,i}(\vb{0}) P^{\vb{E}}_{S,j}(\vb{r})} 
    =
    \frac{r_i r_j}{r^2}
    \Big[ 3\left(R^4-14 R^2+24\right) R \sin (2 R)
     \\
    & +\left(16 R^4-69 R^2+36\right) \cos (2 R) 
    + 2 R^4-3 R^2 -36 
    \Big] \Big/ 8 R^8
    \\
    & +
    \delta_{ij}
    \Big[ \left(-3 R^4+32 R^2-54\right) R \sin (2 R) \\
    & +\left(-13 R^4+52 R^2-27\right) \cos (2 R)
     -R^4+2 R^2+27 
     \Big] \Big/ 8 R^8.
\end{split}
\end{align*}
For each momentum, the (normalized) autocorrelation function of the $x$-component for separation vectors $\vb{r}$ in the $xy$-plane is plotted in the top row of \autoref{fig:2}. 
For the orbital momentum, the degree of correlation is largely positive, and longer-ranged in the longitudinal direction. 
In sharp contrast, components of the spin momentum tend to change sign periodically, and more strongly so in the lateral directions. 
These findings hint at qualitatively distinct spatial organizations for the two currents. 
In the middle row of \autoref{fig:2}, we show 2D streamlines for each momentum in a slice through one realisation of the random field, and colour them according to the value of the $x$-component. 
Zero-crossings of the $x$-component are shown in black to better distinguish regions of ``upwards'' and ``downwards'' flow along $x$. 
We observe that the orbital current keeps the same direction across relatively broad channels, with a typical size in accordance with the correlation function given above. Such structures are channels of energy flow. 
Conversely, the spin current changes direction more frequently, particularly along the lateral ($y$) direction, forming narrow pockets of oppositely directed flow. 
These two contrasting behaviours can seemingly be traced back to the elementary building block of the non-paraxial field, consisting of two interfering plane waves and studied in \cite{bekshaev_2015_transverse}, in which it was found that $\vb{P}_O$ homogeneously points along the bisector, whereas $\vb{P}_S$ oscillates in the transverse direction. 

Corresponding results for the Poynting current, shown in the third column of \autoref{fig:2}, indicate a less clear-cut behaviour, close but not identical to that of the orbital current. 
At this point, it is enlightening to compare the currents of the vector EM field to the simpler case of a random complex scalar field $\Psi = p^\Psi + iq^\Psi$, defined by dropping the polarization term in brackets in \eqref{eq:plane_waves}. 
$\Psi$ obeys the Helmholtz equation with wavevector $k$, and there is a single, divergenceless current $\vb{J} = \frac{1}{2\omega} \Im{\Psi^* \grad \Psi}$. 
Its longitudinal autocorrelation function is given by 
\begin{align*}
    \omega^2 f_J(r) & = \frac{1}{2}
    \left[{C'}^2(r)-C(r)C''(r)\right],
\end{align*}
with $C(r) = L(r) + 2 T(r) = \sin(kr)/kr$ (we remark the similarity of this expression to that for the orbital momentum), and the correlation tensor is
\begin{align*}
\begin{split}
    \expval{J_{i}(\vb{0}) J_{j}(\vb{r})} 
    & =
    \frac{r_i r_j}{r^2}
    \Bigg[
    \frac{\left(2 R^2+R \sin (2 R)+2 \cos (2 R)-2\right)}{4 R^4}
    \Bigg]
    \\
    & +
    \delta_{ij}
    \Bigg[
    -\frac{ (R \sin (2 R)+\cos (2 R)-1)}{4 R^4}
    \Bigg].
\end{split}
\end{align*}
The correlation behaviour of the scalar current, shown in the rightmost column of \autoref{fig:2}, appears to lie in between that of the orbital and Poynting currents, and is similar to both. 
This in turn emphasizes the ``spin'' nature of $\vb{P}_S$, which possesses a behaviour unfound in the scalar case~; it raises the interesting question of how corresponding currents would behave for tensor waves describing other fundamental particles with different spin. 

Finally, we discuss the experimental observability of these optical currents. 
As mentioned in \cite{bekshaev_2011_internal}
, a small probe particle can hardly image subwavelength structures, as its own presence will distort the field on a comparable lengthscale. 
With this in mind, it is tempting to only consider local spatial averages of the currents. 
Our correlation functions suggest that the orbital current will survive local averaging, as it is largely positively correlated to itself over a wavelength-sized volume. 
Conversely, neighbouring pockets of opposite spin flow will cancel each other out. 
In the bottom row of \autoref{fig:2}, we plot the same streamlines again, but after having performed a local average of the field over a spherical volume of diameter $\lambda$ (rendered by the dashed circle). 
The integrated spin current indeed quickly vanishes. 
As a result, orbital and Poynting currents will tend to reconcile, if probed by sufficiently large particles that effectively average over the generic subwavelength inhomogeneities of the spin momentum. 
Consequently, we expect the difference in the orbital and Poynting streamlines highlighted in \autoref{fig:1} to have its significant impact on the motion of very subwavelength objects, such as single atoms or atomic clusters.

\begin{figure}[t!]
    \centering
    \includegraphics[width=\linewidth, trim={0, 0.4cm, 0, 0.6cm}]{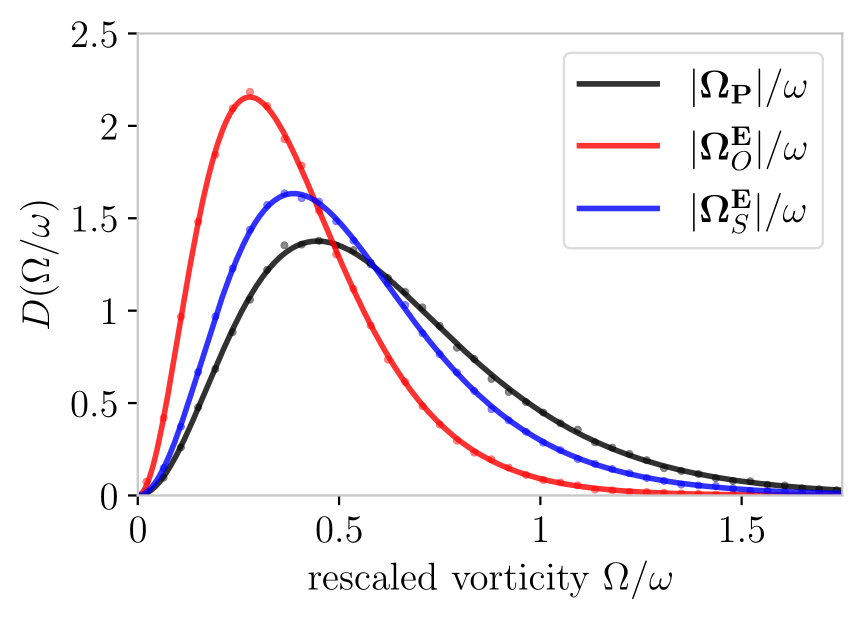}
    \caption{\textbf{Vorticity distributions.} \\
    Analytical (lines) and numerical (dots) probability distributions for the magnitudes of the vorticities of the Poynting, orbital and spin currents. 
    Numerical data was obtained from $10^5$ independent realizations of the random field, each containing $N=10^3$ plane waves.}
    \label{fig:3}
\end{figure}

\subsubsection{Vorticity of the currents}
The tendency of the spin current to ``turn'' more can be further quantified by deriving statistical distributions for the \emph{vorticities} of the optical currents, that were discussed in previous studies \cite{berry_2009_optical, bekshaev_2011_internal}
\begin{align}
    \vb{\Omega_P} = \curl{\vb{P}} 
    \;\;\;\;\;
    \vb{\Omega}^{\vb{E}}_O = \curl{\vb{P}^{\vb{E}}_O} 
    \;\;\;\;\;
    \vb{\Omega}^{\vb{E}}_S = \curl{\vb{P}^{\vb{E}}_S}. 
\end{align}
The strategy for these calculations follows closely that for the magnitudes of the momenta themselves. 
We note that the additional space derivative involved now makes the distributions wavelength-dependent. 
The magnitude distributions for the three vorticities are
\begin{align*}
\begin{split}
    & D(X = \Omega_P/\omega)
    =
    \frac{9 X}{78~886~240} \\ 
    & \cross
    \Big[
     \left(237 \sqrt{5} X +172\right) 819~200
    e^{-6 \sqrt{5} X}
    \\
    & + \left(939~752~400 X-44~642~639 \sqrt{10}\right) \sinh (20 X)
    e^{-8 \sqrt{10} X}
    \\
    & + \left(1~860~213 \sqrt{10} X-880~640\right) 160 \cosh (20 X)
    e^{-8 \sqrt{10} X}
    \Big]
\end{split}
\\[10pt]
\begin{split}
    & D(X = \Omega^{\vb{E}}_O/\omega) =
    \frac{225}{77} X 
    \\ 
    & \cross
    \Big[64 e^{-(15/2) X} -99 e^{-10 X} +35 e^{-6 \sqrt{5} X}\Big]
\end{split}
\\[10pt]
\begin{split}
    & D(X = \Omega^{\vb{E}}_S/\omega) 
    =
    \frac{25 X}{361~504} \\
    & \cross
    \Big[
    83~187 e^{-20 X}
    +9~628~125 e^{-12 X}
    \\
    &
    -5~824~512 e^{-10 X}
    + 286~374 \sqrt{10} \sinh(20X) e^{-8\sqrt{10}X}
    \\
    & 
    -4~792~320 e^{-6 \sqrt{5} X}
    + 905~520 \cosh(20X) e^{-8\sqrt{10}X}
    \Big]
\end{split}
\end{align*}
These distributions are shown in \autoref{fig:3}. 
Despite being identically distributed in magnitude, orbital and spin momenta have different vorticities~: in agreement with the observations of the previous section, that of the spin current is statistically larger. 
An interesting extension of this investigation could be to explore whether or not this relates to some difference in the density of singularities in the orbital and spin flows \cite{bekshaev_2011_internal}. 
The geometry of these singularities, in the special case where all components of the complex electric field vanish, was recently studied in \cite{vernon_2023_3d}, where it was found that the orbital momentum always arranges in elongated ``pseudo vortex lines" in the vicinity of such zeros. 
Visual exploration of the random fields (not shown) indicates that such a coiling structure seems to occur frequently near generic zeros of both the orbital and spin momenta.

\subsection{Democratic momenta and fields with pure helicity}
\label{subsec:democratic}

\begin{figure}[t!]
    \centering
    \includegraphics[width=\linewidth, trim={0, 0.4cm, 0, 0}]{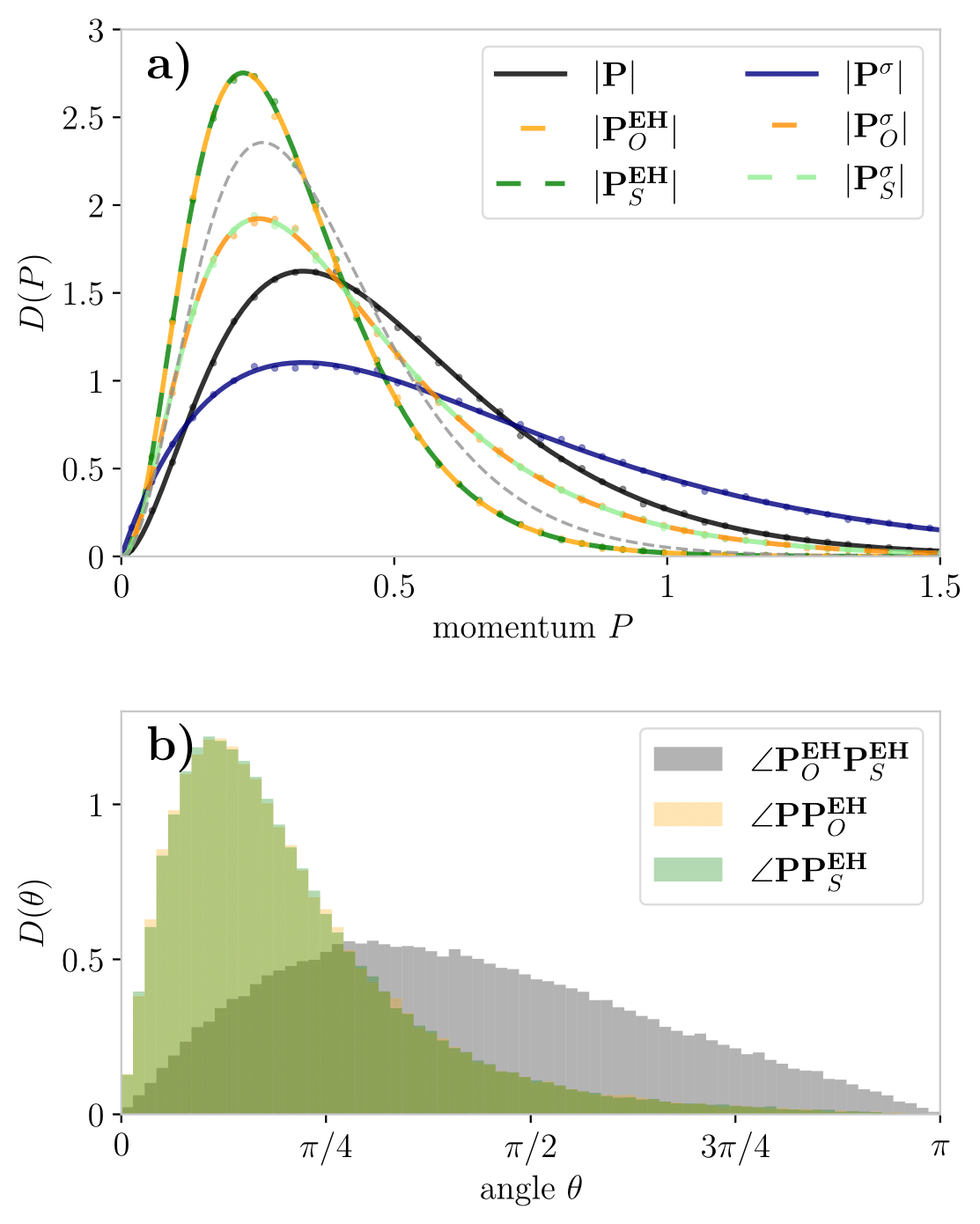}
    \caption{\textbf{Statistics of democratic momenta in randomly polarized and pure helicity fields.} \\
    (a) Analytical (lines) and numerical (dots) probability distributions for the magnitudes of the Poynting, orbital and spin (democratic) momenta in randomly polarized fields (the thin dashed curve shows the distribution for the biased momenta of \autoref{fig:1}), and in pure helicity fields ($\vb{P}^\sigma$). 
    (b) Distribution of the angles between democratic momenta in randomly polarized fields, obtained numerically. Numerical data in this figure was obtained from $10^5$ realizations of the random field, each containing $N=10^3$ plane waves. }
    \label{fig:4}
\end{figure}

Throughout this work, we have focused on electric-biased momenta. 
Equivalent statistics would evidently hold for the magnetic-biased quantities, but not for democratic ones. 
Berry and Shukla recently investigated the difference between biased and democratic quantities in similar statistical calculations \cite{berry_2019_geometry}, and concluded that as a rule of thumb, democratic quantities tend to vary more smoothly and follow narrower distributions. 
Indeed, including contributions from both the electric and magnetic fields (which are uncorrelated to some extent) effectively suppresses regions of strong interference, similarly to the way vector quantities built from three field components also show less interference detail than corresponding scalar quantities. 
We derived the magnitude distributions for the democratic momenta (see SI) and present them in \autoref{fig:4}.a). 
The distribution is still identical for the orbital and spin parts, but is indeed slightly narrower than for the biased momenta (dashed grey line). 
Interestingly, when computing the angle distributions numerically in \autoref{fig:4}.b), we find that the angle between $\vb{P}^{\vb{EH}}_O$ and $\vb{P}^{\vb{EH}}_S$ is on average narrower than that between the corresponding biased quantities. 
As a result, democratic momenta are (slightly) more closely aligned with the Poynting vector than their biased counterparts. 
Our investigations in randomly polarized fields did not reveal more striking differences between biased and democratic momenta, and we believe all qualitative descriptions given in previous sections to hold for democratic currents as well. 

It is enlightening at this point to backtrack on our assumption of \emph{randomly polarized} plane wave components, to consider instead random fields with pure helicity $\sigma=\pm1$. 
This amounts to fixing $\beta_n$ to $0$ or $\pi$ in \eqref{eq:plane_waves}, and enforces $\vb{H} = -i \sigma \vb{E}$ such that biased and democratic quantities become equal (we denote them by a $\sigma$ superscript). 
As detailed in the SI, this adds new non-zero correlations between variables in our statistics, though values of local averages that were already non-zero in the randomly polarized case are unaffected. 
Taking these new correlations into account, we can proceed through similar calculations. 
It is however easy to predict what the distributions will be, as democratic momenta always split into two independent terms originating from components of opposite helicity $\vb{P} = [\vb{P}^+ + \vb{P}^-]/2$ \cite{aiello_2015_note}. 
In a randomly polarized field, this becomes a sum of two independent identically distributed variables, whose distribution simply results from the self-convolution of the distribution for a pure helicity term. 
This easily appears considering the Fourier transform form of our calculations (see SI). 
Distributions in pure helicity fields are also shown and checked against numerics in \autoref{fig:4}.a), and they are broader than all distributions in the randomly polarized case. 
This is likely explained by a weaker ``suppression of interference'' effect, since there is now even less independence between the different components of the EM field.

Finally, it was recently shown by Aiello that for \emph{instantaneous} (that is, not time-averaged) democratic quantities, the fast-oscillating double-frequency terms also happen to be the cross-helicity terms \cite{aiello_2022_one, aiello_2022_helicity}. 
Consequently, cycle-averaging becomes equivalent to ignoring cross-helicity terms, and has therefore no effect on democratic quantities in pure helicity fields. 
For this reason, the distributions derived here for pure helicity fields are also expected to be the magnitude distributions for \emph{instantaneous} democratic momenta (for which the nature of the polarization should be irrelevant). 
Extending our approach to general time-dependent polychromatic fields is beyond the scope of this article, but represents an intriguing avenue, that could highlight profound relations between electric-magnetic democracy, helicity and time-averaging.

\section*{Concluding remarks}

We have investigated various statistical properties of the Poynting, orbital and spin optical momenta in generic isotropic random light fields. 
This isotropy, equivalent to extreme non-paraxiality, was found to increase the discrepancy between Poynting and orbital flows, as the spin momentum unexpectedly becomes equivalent in magnitude to the orbital one. 
Deriving correlation functions, we were able to describe the distinct spatial structures of the orbital and spin currents, the former arranging in broad channels of energy flow akin to those found in a scalar random field, when the latter has higher vorticity and changes direction on a subwavelength scale. 
Upon local averaging over a wavelength-sized volume, the spin current rapidly averages out, leading the orbital and Poynting currents to reconcile. 
Still, the very different behaviour of the orbital and spin currents interrogates what our approach would reveal in other types of waves. 
Indeed, the field-theoretic formalism decomposing the kinetic (Poynting) momentum into canonical (orbital) and Belinfante (spin) parts is of broader generality, and these investigations could be extended and compared to waves describing other particles, such as electrons described by the Dirac equation whose current decomposition into orbital and spin contributions is known as the Gordon decomposition \cite{strange1998relativistic,bliokh_2017_position}, but also to turbulence in acoustic \cite{burns_2020_acoustic} and gravity water waves \cite{bliokh_2022_field}, the latter extensions appearing very natural considering that results from fluid dynamics were used in the present study. 
The spin \emph{angular} momentum density of all types of waves could also be studied, as it is arguably the more relevant quantity from a field-theory perspective, the Belinfante momentum being constructed from it. 

Preliminary investigations have been extended to the random paraxial case (equivalent to the models in \cite{dennis_2002_polarizationc, dennis_2007_nodal}).
In particular, it appears that in analogy with the isotropic case, here the \emph{transverse} orbital and spin momenta have equivalent probability distributions~!
The question about correlations appears more subtle, and depends on the chosen paraxial spectrum~; upon spatial averaging we anticipate this will lead again to dominance of the orbital current. 
Furthermore, such correlations between the orbital and spin parts might indicate features of optical spin-orbit relations distinct from those in the ordered, structured paraxial beams previously studied, although such a discussion goes beyond the scope of this work.

Further investigations of the isotropic electromagnetic case could characterize the generic singularities of the optical currents (isolated points in 3D space) and the statistical geometry of the flows around them, something that has so far only been explored for non-generic zeros of the full complex electric field \cite{vernon_2023_3d}. 
More advanced correlation functions (involving more than two positions, evaluated near extrema, etc\ldots) could reveal finer features of the optical currents as well~; random fields generally offer endless possibilities of statistical investigation \cite{freund_1998_1001}. 

Finally, there appears to be profound links to uncover in relating electric-magnetic democracy, helicity and time-averaging. 
This prompts the extension of our approach to general time-dependent fields, which could require introducing the vector potentials for defining instantaneous momenta, and the weighing of plane wave components by a power spectrum \cite{berry_2000_phase, dennis_2007_nodal}. 
This could represent a step towards better understanding of the spin-orbit decomposition of optical momentum, which as of today remains largely confined to the monochromatic case.

\section*{Data availability statement}

See the Supplementary Information for details on the parameterization of the random fields, tables of elementary averages and correlation functions, general strategies for our computations, and detailed derivations.

\section*{Acknowledgements}

We are grateful to Luke Hands, Nikitas Papasimakis, Michael Berry and Konstantin Bliokh for helpful discussions.
MRD acknowledges support from the EPSRC Centre for Doctoral Training in Topological Design (EP/S02297X/1).
The École Normale Supérieure (Paris) is gratefully acknowledged for TG's fellowship.


\bibliography{References}

\end{document}


\title{Optical momentum distributions in monochromatic,
isotropic random vector fields: \\
Supplementary Information}

\author{Titouan \textsc{Gadeyne}}
\affiliation{School of Physics and Astronomy, University of Birmingham, Birmingham, B15 2TT, UK}
\affiliation{Département de Chimie, École Normale Supérieure, PSL University, 75005 Paris, France}

\author{Mark R. \textsc{Dennis}}
\email{m.r.dennis@bham.ac.uk}
\affiliation{School of Physics and Astronomy, University of Birmingham, Birmingham, B15 2TT, UK}
\affiliation{EPSRC Centre for Doctoral Training in Topological Design, University of Birmingham, Birmingham, B15 2TT, UK}

\maketitle

In this supplementary document, we detail the following aspects of the main text:
\begin{itemize}
    \item Parameterization of the random field. Explicit formulae for computing elementary averages are given and an alternative parameterization is discussed.
    
    \item Gaussian statistics. Elementary variances and correlations are tabulated for fields with random polarization or pure helicity, and computation strategies are outlined.
    
    \item Derivations for all analytical results of the main text. We hope the amount of detail provided can be helpful to students interested in performing similar calculations.
\end{itemize}

\section{Parameterization of the random field}

In this section we give explicit expressions for the random field, required for computing elementary ensemble averages. We briefly compare the intuitive parameterization given in the main text to the one used in the recent paper by Berry and Shukla \cite{berry_2019_geometry}, which has practical advantages. \\

In the main text we give the following parameterization of the random electromagnetic field
\begin{align}
\begin{split}
    \vb{E} & = \begin{aligned}[t]
    \sqrt{\frac{2}{N}} \sum_{n=1}^N
    e^{i \vb{k}_n \vdot \vb{r} + i \psi_n}
    \Big[ 
    e^{i\alpha_n/2} \cos\frac{\beta_n}{2} \vb{e}_{+}(\vb{k}_n)
    + e^{-i \alpha_n/2} \sin\frac{\beta_n}{2} \vb{e}_{-}(\vb{k}_n)
    \Big]
\end{aligned}
\end{split}
\label{eq:plane_waves}
    \\
\begin{split}
    \vb{H} & = \begin{aligned}[t]
    \sqrt{\frac{2}{N}} \sum_{n=1}^N
    e^{i \vb{k}_n \vdot \vb{r} + i \psi_n}
    (-i)
    \Big[  
    e^{i\alpha_n/2} \cos\frac{\beta_n}{2} \vb{e}_{+}(\vb{k}_n)
    - e^{-i \alpha_n/2} \sin\frac{\beta_n}{2} \vb{e}_{-}(\vb{k}_n)
    \Big]
\end{aligned}
\end{split}
\end{align}
where the sum runs over the $N$ plane wave components. The chosen normalization fixes the average electric intensity $\expval{ \Re{|\vb{E}|^2}/2 } = 1$.\\
$\vb{k}_n$ is the wavevector, parameterized by polar and azimuthal angles $(\theta_n, \phi_n)$ on the sphere of directions. In the laboratory Cartesian basis \{$\vb{e}_x, \vb{e}_y, \vb{e}_z$\} it writes
\begin{align*}
    \vb{k}_{n}  = k
    \begin{pmatrix}
    \sin\theta_n \cos\phi_n \\
    \sin\theta_n \sin\phi_n \\
    \cos\theta_n 
    \end{pmatrix}
\end{align*}
$\psi_n$ is the global phase of the plane wave. $\alpha_n$ and $\beta_n$ are respectively the azimuthal and polar angles on the usual Poincaré sphere of polarization, with poles corresponding to circularly polarized states represented by the complex helicity basis vectors $\vb{e}_\sigma(\vb{k}_n)$ transverse to $\vb{k}_n$ (with $\sigma=\pm1$). There are different ways to parameterize this basis, one simple possibility being
\begin{align*}
    \vb{e}_\sigma = \frac{\vb{e}_\theta + i \sigma \vb{e}_\phi }{\sqrt{2}} 
    \;\;\; \text{with} \;\;\;
    \vb{e}_\phi = \frac{\vb{e}_z \cross(\vb{k}/k)}{|\vb{e}_z \cross(\vb{k}/k)|}
    =
    \begin{pmatrix}
        -\sin\phi \\
        \cos\phi \\
        0
    \end{pmatrix}
    , 
    \;\;\;
    \vb{e}_\theta = \vb{e}_\phi \cross(\vb{k}/k)
    =
    \begin{pmatrix}
         \cos\theta \cos\phi \\
         \cos\theta \sin\phi \\
        -\sin\theta
    \end{pmatrix}
\label{eq:helicity_basis}
\end{align*}
Circularly-polarized plane waves are eigenfunctions of the curl operator, $\curl{ (e^{i \vb{k} \vdot \vb{r} } \vb{e}_\sigma )} 
    = \sigma k (e^{i \vb{k} \vdot \vb{r} } \vb{e}_\sigma)$
and we have for a circularly polarized plane wave component $\vb{H}_\sigma(\vb{k}) = -i \sigma \vb{E}_\sigma(\vb{k})$, which can be used to check that the form \eqref{eq:plane_waves} is consistent with Maxwell's equations. \\

In their recent study \cite{berry_2019_geometry}, Berry and Shukla used an alternative parameterization of the same field, using orthogonal \emph{real} basis vectors $\vb{e}_1$ and $\vb{e}_2$
\begin{align*}
    & 
     \vb{e}_\sigma = -i\sigma e^{i\sigma \varsigma} \frac{\vb{e}_1 + i \sigma \vb{e}_2 }{\sqrt{2}}, 
     \;\;\;
     \vb{e}_{1}(\varsigma) = \sin\varsigma \vb{e}_\theta - \cos\varsigma \vb{e}_\phi, 
    \;\;\;
    \vb{e}_{2}(\varsigma)  = \cos\varsigma \vb{e}_\theta + \sin\varsigma \vb{e}_\phi
\end{align*}
Varying the parameter $\varsigma$ simply amounts to rotating the two real vectors around the axis of propagation. They express the random field in the form
\begin{align}
    \vb{E} & = \sqrt{\frac{2}{N}} \sum_{n=1}^N
    e^{i \vb{k}_n \vdot \vb{r}}
    \left[ 
    e^{i\gamma_n} \cos\frac{\mu_n}{2} \vb{e}_{1}(\vb{k}_n, \varsigma_n)
    +
    e^{-i \gamma_n} \sin\frac{\mu_n}{2} \vb{e}_{2}(\vb{k}_n, \varsigma_n)
    \right]
    \\
    \vb{H} & = \sqrt{\frac{2}{N}} \sum_{n=1}^N
    e^{i \vb{k}_n \vdot \vb{r}}
    \left[ 
    e^{i\gamma_n} \cos\frac{\mu_n}{2} \vb{e}_{2}(\vb{k}_n, \varsigma_n)
    -
    e^{-i \gamma_n} \sin\frac{\mu_n}{2} \vb{e}_{1}(\vb{k}_n, \varsigma_n)
    \right]
\label{eq:random_field_2}
\end{align}
which also encompasses all polarizations states and phases. An advantage of using this parameterization is that the real and imaginary parts ($\vb{p}$ and $\vb{q}$) of the complex fields have simpler expressions than when using complex basis vectors. This can be helpful for checking integrals by hand, or when using a symbolic calculation software. \\
It should however be made clear that the angles $\gamma_n$ and $\mu_n$ are no longer the usual angles on the Poincaré sphere. In fact, they correspond to angles on a rotated sphere~: it can be checked that $\mu_n$ is a polar angle defined with respect to the diameter having the orthogonal linear polarization states $\vb{e}_1$ and $\vb{e}_2$ at its ends, and $2\gamma_n$ is an azimuthal angle around this axis. The remaining degree of freedom is spanned by $\varsigma_n$, which affects both the phase and the orientation of the polarization ellipse. The physical meaning of these angles is arguably less intuitive than with the other parameterization, for which $\alpha_n$ rotates the polarization ellipse (thus participating in implementing isotropy), $\psi_n$ is only a phase factor and $\beta_n$ alone controls the ellipticity. Nevertheless, ensemble averaging can equivalently be defined as follows
\begin{align}
    \expval{\bullet}
    = \begin{aligned}[t]
    \prod_{n=1}^N 
    \Bigg[
    \frac{1}{32 \pi^3}
    \int_0^\pi \sin\theta_n \dd \theta_n
    \int_0^{2\pi} \dd \phi_n
    \int_0^\pi \sin\mu_n \dd \mu_n
    \int_0^{\pi} 2 \dd \gamma_n
    \int_0^{2\pi} \dd \varsigma_n
    \Bigg]
    \bullet
    \end{aligned}
\label{eq:averaging2}
\end{align}
The corresponding definition in \cite{berry_2019_geometry} (Eq. 6.8) appears to be missing a factor 2, leading to the peculiar normalization $\expval{1} = 1/2$. This likely is a typographical error, since all averages given in the cited paper are actually consistent with the formula above and with numerical simulations.

\newpage
\section{Gaussian statistics}

\subsection{Table of local variances and correlations}

As explained in the main text, all components of the real and imaginary fields (as well as their derivatives) are centered Gaussian random variables. Ensemble averages become integrals of the form 
\begin{align}
    \expval{\bullet}
    =
    \sqrt{\frac{\det{\vb{\Sigma^{-1}}}}{(2\pi)^M}}
    \int \ldots \int \exp{-\frac{1}{2} \vb{u}^\intercal \vb{\Sigma^{-1}} \vb{u} }
    \bullet \dd^M \vb{u}
\label{eq:Gauss_averaging}
\end{align}
over the vector $\vb{u}$ of all Gaussian variables involved in the quantity $\bullet$. All we require is $\vb{\Sigma}$, the symmetric covariance matrix, with $\Sigma_{ij} = \expval{u_i u_j}$. For local averages, all $u_i$ are evaluated at the same point in space (which can be set to $\vb{r} = \vb{0}$ owing to stationarity). For the isotropic, randomly polarized field, all variances and non-zero correlations involving field components and first spatial derivatives are listed in \autoref{tab:1}. They are consistent with those given by Berry and Shukla \cite{berry_2019_geometry}, however we do include a correlation between derivatives of the electric field that was not mentioned by them.

\begin{table}[h!]
    \centering
    \arraycolsep=2.0pt\def\arraystretch{2.2}
    $\begin{array}{|c|c|c|c|}
    \hline\hline
    \text{Average} & \text{Value} & \text{Conditions} & \text{Symbol}
    \\
    \hline\hline
    \expval{(p^E_i)^2} 
    & 1/3 & \text{ same for $p \leftrightarrow q$ or $E \leftrightarrow H$}
    & \sigma_{x}^2
    \\
    \expval{(\partial_i p^E_i)^2/\omega^2} 
    & 1/15 & \text{ same for $p \leftrightarrow q$ or $E \leftrightarrow H$} 
    & \sigma_{xx}^2
    \\
    \expval{(\partial_i p^E_j)^2/\omega^2} 
    & 2/15 & \text{ $i \neq j$, same for $p \leftrightarrow q$ or $E \leftrightarrow H$} 
    & \sigma_{xy}^2
    \\
    \expval{(p_i^E)(\partial_j q_k^H)/\omega} 
    &
    (-1/6) \epsilon_{ijk}
    & \text{ changes sign when exchanging $p \leftrightarrow q$ or $E \leftrightarrow H$}
    & C
    \\
    \expval{(\partial_ip_j^E)(\partial_k p_l^E)/\omega^2} 
    & 
     (-1/30)[\delta_{jk}\delta_{il}
    +
    \delta_{ij}\delta_{kl}] 
    & \text{ $(i,j) \neq (k,l)$, same for $p \leftrightarrow q$ or $E \leftrightarrow H$}
    & D
    \\ \hline
    \end{array}$
    \caption{Variances and correlations involving field components and their first derivatives, for an isotropic, randomly polarized monochromatic field with frequency $\omega$. Averages involving spatial derivatives are rescaled by $1/\omega$ to produce frequency-independent values. Other averages of pairs of variables are zero (in the randomly polarized case).}
    \label{tab:1}
\end{table}

These values are easily obtained by direct integration over random angles in the plane wave representation of \eqref{eq:plane_waves}, and it is enlightening to actually proceed through some of them by hand, to understand what these values become in isotropic fields with pure helicity. All variables can be put in the form 
\begin{align*}
    u = \sqrt{\frac{2}{N}} \sum_{n=1}^N \Re{ e^{i\psi_n} U_n }
\end{align*}
and averaging over the global phases $\psi_n$ has the effect of removing cross-terms between different plane waves
\begin{align*}
    \prod_{n=1}^N \left[\frac{1}{2\pi}\int_0^{2\pi} \dd \psi_n\right] uv
    & =
    \frac{2}{N} N 
    \frac{1}{2\pi} \int_0^{2\pi} \dd \psi [\cos^2\psi \Re{U}\Re{V} + \sin^2\psi \Im{U}\Im{V}]
    \\
    & =
    \Re{U}\Re{V} + \Im{U}\Im{V}
\end{align*}
where we dropped the index since we only have to average angles for a single plane wave from now on.
$U$ is always proportional to $[e^{i\alpha/2} \cos(\beta/2) e_u \pm e^{-i\alpha/2} \sin(\beta/2) e_u^*]$ with an appropriate complex scalar $e_u$, such that 
\begin{align*}
    \Re{U} & \propto \left[ \cos\frac{\alpha}{2} \Re{e_u}
    - \sin\frac{\alpha}{2}\Im{e_u}
    \right]
    \left[\cos\frac{\beta}{2}
    \pm \sin\frac{\beta}{2}
    \right]
    \\
    \Im{U} & \propto \left[ \sin\frac{\alpha}{2} \Re{e_u}
    + \cos\frac{\alpha}{2}\Im{e_u}
    \right]
    \left[\cos\frac{\beta}{2}
    \mp \sin\frac{\beta}{2}
    \right]
    \\
    \frac{1}{2\pi}\int_0^{2\pi} \dd \alpha
    [\Re{U} \Re{V} + \Im{U} \Im{V}]
    & \propto 
    \left[1
    \pm \cancel{ 2 \cos\frac{\beta}{2} \sin\frac{\beta}{2}}
    +
    1
    \mp \cancel{ 2 \cos\frac{\beta}{2} \sin\frac{\beta}{2}}
    \right]
    \\
    \;\;\; \text{or} \;\;\;
    & \propto
    \left[\cos^2\frac{\beta}{2}
    -
    \sin^2\frac{\beta}{2}
    \right]
\end{align*}
The first situation arises in all averages in \autoref{tab:1}. It is clear that the value of these averages is actually independent of $\beta$ due to the cancellation, and will therefore remains unchanged in the particular case of a field with pure helicity. \\
The second situation clearly depends on whether $\beta$ is integrated over (giving zero) or fixed to a single value (giving $\pm 1$ for pure helicity fields). In the latter case, new non-zero correlations therefore appear, which can be constructed by considering the replacements $\vb{p^E} = \pm \vb{q^H}$ and $\vb{q^E} = \mp \vb{p^H}$. These should then be considered in statistical calculations for pure helicity fields.

\subsection{Computing statistical distributions}

We now review strategies for computing statistical distributions. The probability distribution of a quantity $A(\vb{u})$ is expressed as
\begin{align*}
    D(x) = \expval{\delta(x - A(\vb{u}))}
    =
    \int \frac{\dd s}{2\pi} \exp{-isx} \expval{\exp{i s A(\vb{u})}}
\end{align*}
where we used a Fourier representation of the delta. The average $\expval{\exp{i s A(\vb{u})}}$ (known as the \emph{characteristic function} of the variable $A$) is easily evaluated if $A$ is quadratic in the variables $\vb{u}$, in which cases the following Gaussian integration formula is used
\begin{align}
    \expval{\exp{-\frac{1}{2}\vb{u}^\intercal \vb{A} \vb{u}}}
    & =
    \sqrt{\frac{\det{\vb{\Sigma^{-1}}}}{\det{\vb{\Sigma^{-1} + \vb{A}}}}}
    \;\;\;\;\; 
    \text{with $\vb{A}$ a $D \cross D$ symmetric matrix}
    \label{eq:Gaussian_formula}
\end{align}
it should be noted that, since many of our random variables are uncorrelated and obey the same statistics, the average $\expval{\exp{i s A(\vb{u})}}$ will oftentimes be factorizable as a product of averages over smaller sets of random variables.\\
The remaining integral over $s$ can be evaluated by standard contour integration techniques (Jordan's lemma, residue theorem). 

\paragraph{Exploiting isotropy}
As explained in \cite{berry_2019_geometry}, the probability distribution $D(\abs{\vb{V}})$ of the norm of a three-dimensional, isotropically-distributed vector can be obtained by the sole knowledge of the distribution of one of its components $D_{i}(V_i)$, using the tomography formula
\begin{align}
    D(V) = - 2 V \pdv{D_{i}(V_i)}{V_i} \Big|_{V_i = V}
\label{eq:projection}
\end{align}
Since components of the momenta are already quadratic in the field variables, their magnitudes involve products of four variables, leading to non-Gaussian integrals. Therefore, it is very advantageous to exploit isotropy by computing $D_{i}$ first, which will involve Gaussian integrals only. Isotropy also enforces that $D_i(V_i)$ be even, which allows to perform the contour integrations only for the case $ V_i \geq 0$.

\newpage

\subsection{Computing spatial correlation functions}

We now consider averages involving variables evaluated at two different points in space. We tabulate elementary spatial correlation functions. In the limit $r \rightarrow 0$, these functions can be used to recover the averages of \autoref{tab:1}.

\paragraph{Stationarity}
An important property of our statistics is \emph{stationarity}: averages of two quantities are independent of their absolute positions, and only depend on the separation vector $\vb{r}$ between the two
\begin{align*}
    \expval{u_1(\vb{0}) u_2(\vb{r})} = \expval{u_2(\vb{0}) u_1(-\vb{r})}
\end{align*}

\paragraph{Correlators involving derivatives}
Making use of the fact that the order of derivation and averaging can be interchanged, correlators involving space derivatives can expressed as derivatives of known correlation functions \cite{batchelor_1953_theory, berry_2000_phase}. Writing $G(\vb{r} = \vb{r_2} - \vb{r_1}) = \expval{u(\vb{r_1}) v(\vb{r_2})}$, we have
\begin{align*}
    \expval{u(\vb{r_1}) (\partial_i v)(\vb{r_2})}
    & =
    \expval{u(\vb{r_1}) \partial_{r_{2,i}} v(\vb{r_2})}
    =
    \partial_{r_{2,i}}
    \expval{u(\vb{r_1}) v(\vb{r_2})}
    =
    \partial_{i}
    G(\vb{r})
    \\
    \expval{(\partial_i u)(\vb{r_1}) v(\vb{r_2})}
    & =
    \expval{\partial_{r_{1,i}} u(\vb{r_1}) v(\vb{r_2})}
    =
    \partial_{r_{1,i}}
    \expval{u(\vb{r_1}) v(\vb{r_2})}
    =
    - \partial_{i}
    G(\vb{r})
\end{align*}
An important observation is the introduction of a minus sign for derivatives of quantities at the reference position $\vb{r_1}$.

\paragraph{Two-point correlation functions of EM field components}

The real and imaginary parts of both the electric and magnetic fields are isotropic and divergenceless. The full correlation tensor for the three components of a given field is entirely given by the longitudinal or lateral correlation functions, that we respectively denote $L$ and $T$. They are 
\begin{align*}
    L(r) & = \expval{p^E_x(\vb{0})p^E_x(r \vb{e_x})}
    =
    \frac{\sin(kr) - kr \cos(kr)}{(kr)^3}
    \\
    T(r) & = \expval{p^E_x(\vb{0})p^E_x(r \vb{e_y})}
    =
    \frac{kr \cos(kr) - (1 - (kr)^2) \sin(kr)}{2 (kr)^3}
    \\
    \expval{p^E_i(\vb{0}) p^E_j(\vb{r})}
    & =
    [L(r)-T(r)] \frac{r_i r_j}{r^2} + T(r) \delta_{ij}
\end{align*}
and can be obtained by direct integration of the plane-wave representation of the variables, or from known results of isotropic turbulence theory (\textit{e.g.} 3.4.16 in \cite{batchelor_1953_theory}). These results evidently hold for $\vb{q^E}$, $\vb{p^H}$ and $\vb{q^H}$. Due to the divergenceless nature of these fields, we have
\begin{align*}
    L'(r) = 2\frac{T(r) - L(r)}{r}
\end{align*}
We also note the following useful relations
\begin{align*}
    T'(r) &= - \frac{T(r)}{r} + \frac{L(r)}{r} \left( 1 - \frac{(kr)^2}{2} \right)
    \\
    L''(r) &= 8\frac{L(r) - T(r)}{r^2} - k^2 L(r)
    \\
    T''(r) &= \frac{T(r) - L(r)}{r^2} \left( 4 - (kr)^2 \right)
\end{align*}
We now turn to correlations between components of different fields. Direct integration reveals that the following correlations vanish:
\begin{align*}
    \expval{p^E_i(\vb{0}) q^E_j(\vb{r})} = 0
    \; \; \;
    \expval{p^E_i(\vb{0}) p^H_j(\vb{r})} = 0
\end{align*}
However, $\vb{p^E}$ and $\vb{q^H}$ are correlated, due to Maxwell's equations. Direct integration over angles gives
\begin{align*}
    \expval{p^E_i(\vb{0}) q^H_j(\vb{r})}
    & =
     \frac{1}{2} L(r)
    \sum_k \epsilon_{ijk} k r_k 
\end{align*}
This correlator changes sign when swapping $i$ and $j$, $p$ and $q$, or $E$ and $H$. \\

We note that for the random complex scalar field $\Psi$, the elementary correlation function is
\begin{align*}
    C(r) =  \expval{p^\Psi(\vb{0}) p^\Psi(\vb{r})} = L(r) + 2 T(r) = \frac{\sin(kr)}{kr}
\end{align*}

Finally, we tabulate some correlators involving derivatives of field components, that will be encountered in our calculations. The first one is
\begin{align*}
    \expval{
    p^E_j (\vb{0}) [\partial_k p^E_x] (\vb{R})
    }
    & =
    \partial_k 
    \expval{
    p^E_j (\vb{0}) p^E_x (\vb{r})
    } \Bigg|_{\vb{r} = \vb{R}}
    \\
    & =
    \partial_k \left[
    \frac{L(r)- T(r)}{r^2} r_j x + \delta_{xj} T(r)
    \right] \Bigg|_{\vb{r} = \vb{R}}
    \\
    & =
    \left[
    \frac{x r_j r_k}{r^3} \left(L'(r)- T'(r) - 2\frac{L(r)- T(r)}{r}\right)
    +
    \frac{L(r)- T(r)}{r^2}
    \partial_k(r_j x)
    + \delta_{xj} \frac{r_k}{r} T'(r)
    \right] \Bigg|_{\vb{r} = \vb{R}}
    \\
    & =
    \left[
    \frac{x r_j r_k}{r^3} \left[2 L'(r)- T'(r)\right]
    -
    \frac{L'(r)}{2r}
    [x \delta_{kj} + r_j \delta_{xk} ]
    + \delta_{xj} \frac{r_k}{r} T'(r)
    \right] \Bigg|_{\vb{r} = \vb{R}}
\end{align*}
The case of interest will be $\vb{R} = (r, 0, 0)$, simplifying to
\begin{align}
\begin{split}
    \expval{
    p^E_j (\vb{0}) [\partial_k p^E_x] (r \vb{e}_x)
    }
    & =
    \left[ \delta_{xj} \delta_{xk}
    \left[2 L'(r)- T'(r)\right]
    -
    \frac{L'(r)}{2r}
    [r \delta_{kj} + r \delta_{xj} \delta_{xk} ]
    + \delta_{xj} \delta_{xk} T'(r)
    \right]
    \\
    & = L'(r) \Big[
    \frac{3 \delta_{xj} \delta_{xk} }{2}
    -
    \frac{\delta_{kj}}{2}
    \Big]
\end{split}
\end{align}

We turn to the following average, involving two derivatives. We have, exploiting the previous result
\begin{align*}
    \expval{
    [\partial_j p^E_x](\vb{0}) [\partial_j p^E_x]( \vb{R})
    }
    & = - \partial_j
    \expval{
    p^E_x(\vb{0}) [\partial_j p^E_x](\vb{r})
    } \Bigg|_{\vb{r} = \vb{R}}
    \\
    & =
    - \partial_j
    \left[
    \frac{x^2 r_j}{r^3} [2L'(r)- T'(r)]
    -
    \frac{L'(r)}{2r}
    [2 x \delta_{xj}]
    + \frac{r_j}{r} T'(r)
    \right]
    \Bigg|_{\vb{r} = \vb{R}}
    \\
    & =
    -
    \Bigg[
    \frac{x^2 + 2x^2 \delta_{xj}}{r^3} [2L'(r)- T'(r)]
    -
    \frac{3x^2 r_j^2}{r^5} [2L'(r)- T'(r)]
    +
    \frac{x^2 r_j^2}{r^4} [2L''(r)- T''(r)]
    \\
    & -
    \delta_{xj} \frac{L'(r)}{r}
    +
    \delta_{xj} \frac{x^2}{r^3} L'(r)
    -
    \delta_{xj} \frac{x^2}{r^2} L''(r)
    + 
    \frac{T'(r)}{r}
    -
    \frac{r_j^2}{r^3} T'(r)
    +
    \frac{r_j^2}{r^2} T''(r)
    \Bigg]
    \Bigg|_{\vb{r} = \vb{R}}
\end{align*}
Evaluating at $\vb{R} = (r, 0, 0)$, we find
\begin{align}
\begin{split}
    \expval{
    [\partial_j p^E_x](\vb{0}) [\partial_j p^E_x]( r\vb{e}_x)
    }
    & =
    -
    \Bigg[
    \frac{1 + 2 \delta_{xj}}{r} [2L'(r)- T'(r)]
    -
    \frac{3 \delta_{xj}}{r} [2L'(r)- T'(r)]
    +
    \delta_{xj} [2L''(r)- T''(r)]
    \\
    & -
    \delta_{xj} \frac{L'(r)}{r}
    +
    \delta_{xj} \frac{1}{r} L'(r)
    -
    \delta_{xj} L''(r)
    + 
    \frac{T'(r)}{r}
    -
    \frac{\delta_{xj}}{r} T'(r)
    +
    \delta_{xj} T''(r)
    \Bigg]
    \\
    & =
    -
    \Bigg[
    \frac{2 L'(r)}{r}
    +
    \delta_{xj} \left( L''(r) - \frac{2 L'(r)}{r} \right)
    \Bigg]
\end{split}
\end{align}

\newpage
\section{Detailed derivations}

\subsection{Magnitude distributions for the momenta}

\subsubsection*{Poynting momentum}
The time-averaged Poynting momentum writes
\begin{align*}
    \vb{P} = \frac{1}{2} \Re{ \vb{E}^* \cross \vb{H} }
    =
    \frac{1}{2} \left[ \vb{p^E} \cross \vb{p^H} + \vb{q^E} \cross \vb{q^H}  \right]
\end{align*}
Exploiting isotropy, we will first compute the distribution for the $x$-component
\begin{align*}
    D_x(P_x) & = \expval{\delta \Big(P_x - \sum_{j,k} \frac{1}{2} \epsilon_{xjk}\left[ p^E_j p^H_k + q^E_j q^H_k  \right] \Big)}
    =
    \int \frac{\dd s}{2\pi} \exp{-isP_x} \expval{\exp{i s \frac{1}{2} p_y^E p_z^H}}^4
    \\
    & =
    \int \frac{\dd s}{2\pi} \exp{-isP_x} 
    \left[ 
    \frac{1}{1 + s^2 \sigma_x^4 / 4}
    \right]^2
    =
    \frac{1 + 2|P_x|/\sigma_x^2}{2\sigma_x^2}
    \exp{-\frac{2 |P_x|}{\sigma_x^2}} 
\end{align*}
where $\sigma_x^2 = \expval{(p^E_x)^2} = 1/3$ (see tabulated variances). The second step involves factorization of the average using independence of field components (indeed, none of the non-zero correlations of \autoref{tab:1} are involved here), the third step uses \eqref{eq:Gaussian_formula}, and the last step involves integration in the complex plane, using the residue theorem at the second-order poles $s=\pm 2i/\sigma_x^2$ (for many of the derivations to come, residues were obtained using a symbolic calculation software, by finding the relevant term in the Laurent series). The distribution of the norm of the Poynting vector is obtained by use of \eqref{eq:projection}, yielding
\begin{align}
    D(P) & = 
    \frac{4P^2 }{\sigma_x^6}
    \exp{-\frac{2 P}{\sigma_x^2}} 
    =
    108 P^2 \exp{-6 P}
\end{align}
This distribution is independent of $k$, the magnitude of the momentum only scales with the overall intensity of the field. 

\subsubsection*{Orbital and spin momenta (biased)}
The canonical and spin momenta read
\begin{align*}
    \vb{P^E_O} & = \frac{1}{2\omega} \Im{ \vb{E}^* \vdot ( \grad )\vb{E}  } 
    =
    \frac{1}{2\omega} \vb{p^E} \vdot ( \grad) \vb{q^E} 
    -
    \frac{1}{2\omega} 
    \vb{q^E} \vdot ( \grad )\vb{p^E}  
    \\
    \vb{P^E_S} & = -\frac{1}{2\omega} \Im{ (\vb{E}^* \vdot  \grad) \vb{E} ) } 
    =
    -\frac{1}{2\omega} (\vb{p^E} \vdot \grad) \vb{q^E} 
    +
    \frac{1}{2\omega} 
    (\vb{q^E} \vdot \grad) \vb{p^E} 
\end{align*}
It is advantageous at this point to absorb the factor $1/\omega$ in the space derivatives, because the rescaled variances $\sigma_{ij}^2 = \expval{(\partial_i p_j)^2/\omega^2} \propto k^2/\omega^2 = 1$ become constants independent of $k$. We start by computing the distribution for the $x$-component of the orbital momentum
\begin{align*}
    D_x(P^E_{O,x}) & = \expval{\delta \Big(P^E_{O,x} - \sum_{j} \frac{1}{2}\left[ p^E_j \partial_x q^E_j - q^E_j \partial_x p^E_j  \right] \Big)} \\
    & =
     \int \frac{\dd s}{2\pi} \exp{-isP^E_{O,x}} \expval{\exp{i s \frac{1}{2} p^E_x \partial_x q^E_x}}^2
    \expval{\exp{i s \frac{1}{2} p^E_y \partial_x q^E_y}}^4
    \\
    & =
     \int \frac{\dd s}{2\pi} \exp{-isP^E_{O,x}} 
     \left[ 
    \frac{1}{1 + s^2 \sigma_x^2 \sigma_{xx}^2/4}
    \right]
    \left[ 
    \frac{1}{1 + s^2 \sigma_x^2 \sigma_{xy}^2/4}
    \right]^2
    \\
    & =
    \frac{1}{(A-B)^2}
    \left[
    \frac{A^{3/2}}{2} \exp{-\frac{|P^E_{O,x}|}{\sqrt{A}}}
    -
    \frac{\sqrt{B}(3A-B)+(A-B)|P^E_{O,x}|}{4}
    \exp{-\frac{|P^E_{O,x}|}{\sqrt{B}}}
    \right]
\end{align*}
where $A = \sigma_x^2\sigma_{xx}^2/4 = 1/180$ and $B = \sigma_x^2\sigma_{xy}^2/4 = 1/90$. The distribution for the norm is again obtained using \eqref{eq:projection}
\begin{align}
\begin{split}
    D(P^E_{O}) 
    & =
    \frac{P^E_O}{(A-B)^2}
    \left[
    A \exp{-\frac{P^E_{O}}{\sqrt{A}}}
    -
    \frac{2A+(A-B)P^E_{O}/\sqrt{B}
    }{2}
    \exp{-\frac{P^E_{O}}{\sqrt{B}}}
    \right]
    \\
    & =
    180 P^E_{O}
    \left[
    \exp{-6\sqrt{5} P^E_{O}}
    - \left[ 1
    -\frac{3\sqrt{10}}{2}
    P^E_{O}
    \right]
    \exp{-3 \sqrt{10} P^E_{O}}
    \right]
\end{split}
\end{align}
Perhaps surprisingly, we find that in calculating the distribution for the $x$-component of the spin momentum, the averages to compute are the same. Indeed, we have
\begin{align*}
    D_x(P^E_{S,x}) & = \expval{\delta \Big(P^E_{S,x} - \sum_{j} \frac{1}{2}\left[ p^E_j \partial_j q^E_x - q^E_j \partial_j p^E_x  \right] \Big)} \\
    & =
     \int \frac{\dd s}{2\pi} \exp{-isP^E_{S,x}} \expval{\exp{i s \frac{1}{2} p^E_x \partial_x q^E_x}}^2
    \expval{\exp{i s \frac{1}{2} p^E_y \partial_y q^E_x}}^4
\end{align*}
and since $\partial_x q^E_y$ and $\partial_y q^E_x$ are both uncorrelated to $p^E_y$ and have the same variance (see \autoref{tab:1}), the rest of the calculation is identical, and we find that the orbital and spin momenta have the exact same magnitude distribution. We also find that in our random field, their distribution is independent of $k$, like that of the Poynting momentum.

\subsubsection*{Orbital and spin momenta (democratic)}
We repeat the previous calculation, using the democratic definition of the momenta
\begin{align*}
    \vb{P_O^{EH}}
    =
    \frac{1}{4\omega} \Im{
    \vb{E}^* \vdot (\grad) \vb{E} + 
    \vb{H}^* \vdot (\grad) \vb{H}
    }
    =
    \frac{1}{4\omega} \left[
    \vb{p^E} \vdot (\grad) \vb{q^E} - \vb{q^E} \vdot (\grad) \vb{p^E}
    + 
    \vb{p^H} \vdot (\grad) \vb{q^H} - \vb{q^H} \vdot (\grad) \vb{p^H}
    \right]
\end{align*}
The distribution of the $x$-component is
\begin{align*}
    D_x(P^{EH}_{O,x}) & = 
    \expval{\delta \left(
    P^{EH}_{O,x} - \vb{P_O^{EH}} \vdot \vb{e}_x
    \right)}
    \\
    & =
     \int \frac{\dd s}{2\pi} \exp{-i s P^{EH}_{O,x}} 
    \expval{\exp{is \frac{1}{4} 
    p^E_x \partial_x q^E_x 
    }}^4
    \expval{\exp{is \frac{1}{4} \left[
    p^E_y \partial_x q^E_y + p^H_z \partial_x q^H_z
    \right]
    }}^4
    \\
    & =
     \int \frac{\dd s}{2\pi} \exp{-i s P^{EH}_{O,x}} 
    \left[
    \frac{1}{1+s^2\sigma_x^2\sigma_{xx}^2/16}
    \right]^2
    \left[
    \frac{256}{C^4 s^4-2 C^2 s^2 (s^2 \sigma_x^2 \sigma_{xy}^2-16) +\left(s^2 \sigma_x^2 \sigma_{xy}^2+16\right)^2}
    \right]^2
    \\
    & =
     \int \frac{\dd s}{2\pi} \exp{-i s P^{EH}_{O,x}} 
    \left[
    \frac{1}{1+s^2A/4}
    \right]^2
    \left[
    \frac{1}{s^4(C^2 - 4B)^2/256 + s^2 (C^2 + 4 B)/8 +1}
    \right]^2
\end{align*}
with $C= \pm 1/6$ (see \autoref{tab:1}) capturing a correlation between electric and magnetic variables. We find, using numerical values to reduce the length of the formula
\begin{align*}
    D_x(P^{EH}_{O,x}) =
    &\frac{2~304 \left(1~580 \abs{P^{EH}_{O,x}}+141 \sqrt{5}\right)}{493~039}
    \exp{-12 \sqrt{5} \abs{P^{EH}_{O,x}} }
    \\
    & 
    + \frac{9 \exp{-16 \sqrt{10} \abs{P^{EH}_{O,x}}} }{315~544~960}  
    \Bigg[ 32 \left(14~031~980 \sqrt{10} \abs{P^{EH}_{O,x}} +860~029\right) \sinh (40 \abs{P^{EH}_{O,x}})
    \\
    & +5 \left(284~038~496 \abs{P^{EH}_{O,x}} +1~742~263 \sqrt{10}\right) \cosh (40 \abs{P^{EH}_{O,x}})
    \Bigg]
\end{align*}
The magnitude distribution is
\begin{align}
\begin{split}
D(P^{EH}_{O}) = &
9  P^{EH}_{O}
\Bigg[
1~638~400 \left(237 \sqrt{5} P^{EH}_{O} +86\right) \exp{-12 \sqrt{5} P^{EH}_{O}}
\\
& +
\left(1~879~504~800 P^{EH}_{O}-44~642~639 \sqrt{10}\right) \sinh (40 P^{EH}_{O}) \exp{-16 \sqrt{10} P^{EH}_{O}}
\\
& +
320 \left(1~860~213 \sqrt{10} P^{EH}_{O}-440~320\right) \cosh (40 P^{EH}_{O}) \exp{-16 \sqrt{10} P^{EH}_{O}}
\Bigg] 
\Bigg/ 19~721~560
\end{split}
\end{align}
For the same reasons as before, we find very early in the derivation that the distribution of the spin momentum will again be identical to that of the orbital one.

\subsubsection*{Poynting momentum (pure helicity)}
Considering now a random field with pure helicity, we have $\vb{E} = \pm i \vb{H}$, and the time-averaged Poynting momentum simplifies
\begin{align*}
    \vb{P}^\sigma = \frac{1}{2} \Re{ \vb{E}^* \cross \vb{H} }
    =
    \frac{1}{2} \left[ \vb{p^E} \cross \vb{p^H} + \vb{q^E} \cross \vb{q^H}  \right]
    =
    \left[ \vb{p^E} \cross \vb{p^H}  \right]
\end{align*}
The derivation becomes
\begin{align*}
    D_x(P^\sigma_x) & = \expval{\delta \Big(P^\sigma_x - \sum_{j,k} \epsilon_{xjk}p^E_j p^H_k \Big)}
    =
    \int \frac{\dd s}{2\pi} \exp{-isP^\sigma_x} \expval{\exp{i s p_y^E p_z^H}}^2
    \\
    & =
    \int \frac{\dd s}{2\pi} \exp{-isP^\sigma_x} 
    \left[ 
    \frac{1}{1 + s^2 \sigma_x^4}
    \right]
    = \frac{1}{2 \sigma_x^2} \exp{- \abs{P^\sigma_x} / \sigma_x^2 } 
    = \frac{3}{2} \exp{- 3\abs{P^\sigma_x}}
\end{align*}
and the magnitude distribution is
\begin{align}
    D(P^\sigma) & = 
    9 P^\sigma \exp{- 3 P^\sigma } 
\end{align}
Note how this result could have been predicted~: we know $\vb{P}$ naturally splits into independent helicity components $[\vb{P}^+ + \vb{P}^-]/2$, that are identically distributed in the randomly polarized field. And indeed, the integrand in the Fourier transform in the second step of the derivation above is simply the square root of that for the democratic momenta in the randomly polarized case (up to the factor 2 of course), such that the latter is simply a self-convolution of the pure helicity result. \\
Another interesting observation is that this distribution is also the one followed by the \emph{instantaneous} Poynting momentum in the randomly polarized case, since we have $\vb{P}^\sigma = \Pf = \E \cross \Hf$ and none of the new correlations were involved here.

\subsubsection*{Orbital and spin momenta (pure helicity)}

In a pure helicity field, the momenta rewrite
\begin{align*}
    \vb{P_O}^\sigma
    =
    \frac{1}{2\omega} \left[
    \vb{p^E} \vdot (\grad) \vb{q^E} 
    + 
    \vb{p^H} \vdot (\grad) \vb{q^H} 
    \right]
\end{align*}
It can be checked that using this choice of variables, none of the new non-zero correlations will be involved.
The derivation becomes
\begin{align*}
    D_x(P^\sigma_{O,x}) 
    & =
     \int \frac{\dd s}{2\pi} \exp{-i s P^\sigma_{O,x}} 
    \expval{\exp{is \frac{1}{2} 
    p^E_x \partial_x q^E_x 
    }}^2
    \expval{\exp{is \frac{1}{2} \left[
    p^E_y \partial_x q^E_y + p^H_z \partial_x q^H_z
    \right]
    }}^2
    \\
    & =
     \int \frac{\dd s}{2\pi} \exp{-i s P^\sigma_{O,x}} 
    \left[
    \frac{1}{1+s^2A}
    \right]
    \left[
    \frac{1}{s^4(C^2 - 4B)^2/16 + s^2 (C^2 + 4 B)/2 +1}
    \right]
    \\
    & = 
    \frac{3}{632} \Bigg[-128 \sqrt{5} \exp{-6 \sqrt{5} |P^\sigma_{O,x}|} 
     + \left( 214 \sqrt{10} \cosh( 20 |P^\sigma_{O,x}| )
    + 664 \sinh( 20 |P^\sigma_{O,x}| ) \right)
    \exp{-8\sqrt{10}|P^\sigma_{O,x}|}
    \Bigg]
\end{align*}
and the magnitude distribution is
\begin{align}
    D(P^\sigma_{O})
    =
    \frac{9}{79} P^\sigma_{O} 
    \Bigg[
    & -320 \exp{- 6 \sqrt{5} P^\sigma_{O}}
    +
    \left( 
    320 \cosh( 20 P^\sigma_{O} )
    + 86 \sqrt{10} \sinh( 20 P^\sigma_{O} )
    \right)
    \exp{- 8 \sqrt{10} P^\sigma_{O}} 
    \Bigg]
\end{align}
Inspecting the steps of the derivation, it is evident that the self-convolution of this pure helicity result will produce the randomly polarized case (up to the appropriate factor 2). This can again be understood from the splitting $\vb{P_O^{EH}} = [\vb{P_O}^+ + \vb{P_O}^-]/2$. \\
As before, an identical derivation exists for the spin momentum. 

\newpage
\subsection{Vorticity distributions}

\subsubsection*{Poynting vorticity}

We have
\begin{align*}
    \vb{\Omega_P} = \curl{\vb{P}} = 
    \frac{1}{2} \left[ 
    (\vb{p^H} \vdot \grad) \vb{p^E} - (\vb{p^E} \vdot \grad) \vb{p^H}
    +
    (\vb{q^H} \vdot \grad) \vb{q^E} - (\vb{q^E} \vdot \grad) \vb{q^H}
    \right]
\end{align*}
Again, as the vorticity is isotropically distributed, we may derive the distribution of the $x$-component only. Due to the additional space derivative, the distributions will acquire a $k$-dependence, and it will be more convenient to derive distributions of rescaled vorticities $\Omega/\omega$ to keep working with the frequency-independent variances of \autoref{tab:1}. \\
Careful analysis of the non-zero correlations between variables allows to neatly factorize the average
\begin{align*}
    D_x(X = \Omega_{P,x}/\omega) & = \expval{\delta(X - \frac{1}{2} \left[  
    (\vb{p^H} \cdot \nabla) p^E_x 
    -
    (\vb{p^E} \cdot \nabla) p^H_x
    +
    (\vb{q^H} \cdot \nabla) q^E_x
    -
    (\vb{q^E} \cdot \nabla) q^H_x
    \right])} \\
    & =
    \int \frac{\dd s}{2\pi} \exp{-isX} \expval{ \exp{is \frac{1}{2} p^H_x \partial_x p^E_x }}^4
    \expval{ \exp{is \frac{1}{2} [ p^H_y \partial_y p^E_x+
    q^H_z \partial_z q^E_x]}}^4
    \\
    & =
    \int \frac{\dd s}{2\pi} \exp{-isX} 
    \left[ 
    \frac{1}{1 + A s^2}
    \right]^2
    \left[ 
    \frac{1}{s^4 (C^2-4 B)^2/16 + s^2 (C^2+4B)/2 + 1}
    \right]^2
\end{align*}
with constants defined as before. This calculation turns out to be very close to that of the democratic orbital momentum. We have the same integral, when replacing $s \rightarrow s/2$, $X \rightarrow 2 X$ and multiplying everything by 2. Therefore $D_x(X) = \frac{1}{2} D_x(2 P^{EH}_{O,x})$. 
With numerical values, this is
\begin{align*}
    D_x(X = \Omega_{P,x}/\omega) 
    = &
    \Bigg[
    1~474~560 \left(790 \abs{X}+141 \sqrt{5}\right) \exp{-6 \sqrt{5} \abs{X}}
    \\
    + & 288 \left(7~015~990 \sqrt{10} \abs{X}+860~029\right) \sinh (20 \abs{X}) \exp{-8 \sqrt{10} \abs{X}}
    \\
    + & 45 \left(142~019~248 \abs{X}+1~742~263 \sqrt{10}\right) \cosh (20 \abs{X}) \exp{-8 \sqrt{10} \abs{X}}
    \Bigg]  \Bigg/ 631~089~920
\end{align*}
The distribution for the magnitude can be derived from the above results (or directly from that of the democratic orbital momentum) and reads
\begin{align}
\begin{split}
    D(X = \Omega_{P}/\omega) 
    = &
    9 X
    \Bigg[
     \left(237 \sqrt{5} X +172\right) 819~200
    \exp{-6 \sqrt{5} X}
    \\
    + &  \left(939~752~400 X-44~642~639 \sqrt{10}\right) \sinh (20 X)
    \exp{-8 \sqrt{10} X}
    \\
    + &  \left(1~860~213 \sqrt{10} X-880~640\right) 160 \cosh (20 X)
    \exp{-8 \sqrt{10} X}
    \Bigg]
    \Bigg/ 78~886~240
\end{split}
\end{align}

\subsubsection*{Orbital momentum vorticity (biased)}

The vorticity of the orbital momentum writes
\begin{align*}
    \vb{\Omega^E_O} = \curl{ \vb{P^E_O} }
    & =
    \sum_{j,k,l}
    \frac{1}{2\omega} \left[  \epsilon_{ijk} \partial_j ( p^E_l \partial_k q^E_l)
    -
    \epsilon_{ijk} \partial_j ( q^E_l \partial_k p^E_l)
    \right]
    =
    \frac{1}{\omega}
    \sum_{j,k,l}
     \epsilon_{ijk} (\partial_j p^E_l) (\partial_{k} q^E_l)
     \\
    & =
    \frac{1}{\omega}
    (\grad \vb{p^E}) \cross \vdot~(\grad \vb{q^E})
\end{align*}
where we used a notation similar to Berry's in the last step, with the scalar product connecting $\vb{p^E}$ and $\vb{q^E}$ and the cross product connecting the gradient operators \cite{berry_2009_optical}. We note that all second derivatives of the field components cancelled out. The distribution of the $x$-component is
\begin{align*}
    D_x(X = \Omega^E_{O,x}/\omega)
    & =
     \expval{\delta(X - \left[  
    (\partial_y \vb{p^E}) \vdot (\partial_z \vb{q^E})
     -
     (\partial_z \vb{p^E}) \vdot (\partial_y \vb{q^E})
    \right])} 
    \\
    & =
    \int \frac{\dd s}{2\pi} \exp{-isX} \expval{ \exp{is   
    (\partial_y p_x) (\partial_z q_x)
    }}^2
    \expval{\exp{is
    [ (\partial_y p_y) (\partial_z q_y)
     -
     (\partial_z p_z) (\partial_y q_z) ]
    }
    }^2
    \\
    & =
    \int \frac{\dd s}{2\pi} \exp{-isX} 
    \left[
    \frac{1}{1+ s^2 \sigma_{xy}^2}
    \right]
    \left[
    \frac{1}{D^4 s^4-D^2 s^2 \left(s^2 \left(\sigma_{xx}^4+\sigma_{xy}^4\right)+2\right)+\left(\sigma_{xx}^2 \sigma_{xy}^2 s^2+1\right)^2}
    \right]
    \\
    & =
    \int \frac{\dd s}{2\pi} \exp{-isX} 
    \left[
    \frac{1}{1+ s^2 \sigma_{xy}^2}
    \right]
    \left[
    \frac{1}{
    s^4 (D^2 - \sigma_{xx}^4)(D^2 - \sigma_{xy}^4)
    -
    2 s^2 (D^2 - \sigma_{xx}^2 \sigma_{xy}^2)
    +
    1}
    \right]
    \\
    & =
    \frac{1}{4} \Bigg[
    \frac{2 \sigma_{xy}^6}{D^4+D^2 \left(\sigma_{xy}^4-\sigma_{xx}^4\right)+\sigma_{xy}^4 \left(\sigma_{xx}^2-\sigma_{xy}^2\right)^2} 
    \exp{-\frac{\abs{\Omega^E_{O,x}}}{\sigma_{xy}^2}}
    \\
    & +
    \frac{\left[-\left(D+\sigma_{xx}^2\right) \left(D-\sigma_{xy}^2\right)\right]^{3/2}
    }{D (\sigma_{xx}^2-\sigma_{xy}^2)\left[ (D-\sigma_{xy}^2) \left(D+\sigma_{xx}^2\right)+\sigma_{xy}^4\right]}
    \exp{-\frac{\abs{\Omega^E_{O,x}}}{\sqrt{-\left(D+\sigma_{xx}^2\right) \left(D-\sigma_{xy}^2\right)}}}
    \\
    & - \frac{ \left[ -\left(D-\sigma_{xx}^2\right) \left(D+\sigma_{xy}^2\right) \right]^{3/2}
    }{ D \left(\sigma_{xx}^2-\sigma_{xy}^2\right) \left[\left(D-\sigma_{xx}^2\right) \left(D+\sigma_{xy}^2\right) + \sigma_{xy}^4\right]}
    \exp{-\frac{\abs{\Omega^E_{O,x}}}{\sqrt{-\left(D-\sigma_{xx}^2\right) \left(D+\sigma_{xy}^2\right)}}}
    \Bigg]
\end{align*}
with $D = -1/30$ (see tables). This evaluates to
\begin{align*}
    D_x(X = \Omega^E_{O,x}/\omega) 
    = &
    \frac{15}{308} \left( 256 \exp{-(15/2) \abs{X}}-297 \exp{-10 \abs{X}}+35 \sqrt{5} \exp{-6 \sqrt{5} \abs{X}}\right)
\end{align*}
The magnitude distribution is
\begin{align}
    D(X = \Omega^E_O/\omega) 
    = &
    \frac{225}{77} X \left(64 \exp{-(15/2) X} -99\exp{-10 X} +35 \exp{-6 \sqrt{5} X}\right)
\end{align}

\subsubsection*{Spin momentum vorticity (biased)}

The vorticity reads
\begin{align}
    \vb{\Omega^E_S} = \curl{ \vb{P^E_S} }
    & =
    -\sum_{j,k,l}
    \frac{1}{2\omega} \epsilon_{ijk} \left[  \partial_j ( p^E_l \partial_l q^E_k)
    -
     \partial_j ( q^E_l \partial_l p^E_k)
    \right]
    \\
    & =
    -\sum_{j,k,l}
    \frac{1}{2\omega} \epsilon_{ijk} \left[  
    (\partial_j  p^E_l) (\partial_l q^E_k)
    + p^E_l (\partial_j \partial_l q^E_k)
    -
     (\partial_j  q^E_l) (\partial_l p^E_k)
     -
      q^E_l ( \partial_j \partial_l p^E_k)
    \right]
    \\
    & =
    -\sum_{j,k,l}
    \frac{1}{2\omega} \left[  \epsilon_{ijk} (\partial_j p^E_l) (\partial_{l} q^E_k)
    +
    \omega p^E_l (\partial_{l} p^H_i)
    -
    \epsilon_{ijk} (\partial_j q^E_l) (\partial_{l} p^E_k)
    +
    \omega q^E_l (\partial_{l} q^H_i)
    \right]
\end{align}
where in the last step, we used Maxwell's equation $\curl{\vb{E}} = i\omega \vb{H}$ to express second derivatives of the electric variables using first derivatives of magnetic variables. This removes the need to evaluate variances and correlations involving the second derivatives. With some work, the average required for the distribution can be factorized as a product of four simpler averages
\begin{align*}
    & D_x(X = \Omega^E_{S,x}/\omega )
    =
    \expval{\delta\left(X - \vb{\Omega^E_S} \vdot \vb{e}_x\right)} 
    \\
    & = \begin{aligned}[t]
    \int \frac{\dd s}{2\pi} \exp{-isX} 
    \expval{ \exp{i\frac{s}{2}  
    [ \partial_y p^E_x \partial_x q^E_z
    + \partial_x p^E_y \partial_z q^E_x ]
    }}^2
    \expval{ \exp{i\frac{s}{2}  
    [ (\partial_y p^E_z + \partial_z p^E_y)  (\partial_z q^E_z - \partial_y q^E_y) ]
    }}^2
    \\
    \cross 
    \expval{\exp{i\frac{s}{2}
    [p^E_x \partial_x p^H_x]
    }}^2
    \expval{\exp{i\frac{s}{2}
    [p^E_y \partial_y p^H_x
    +q^E_z \partial_z q^H_x]
    }}^2
    \end{aligned}
    \\
    & = \begin{aligned}[t]
    \int \frac{\dd s}{2\pi} \exp{-isX} 
    \left[
    \frac{1}{s^4 \left(D^2-\sigma_{xy}^4\right)^2/16+ s^2 \left(D^2+\sigma_{xy}^4\right)/2+1}
    \right] \left[
    \frac{1}{1 + t^2(\sigma_{xx}^2 - D)(\sigma_{xy}^2 + D)}
    \right]
    \\
    \cross
    \left[
    \frac{1}{1 + s^2 A}
    \right] \left[
    \frac{1}{s^4 \left(C^2-4B\right)^2/16 +  s^2 \left(C^2+4B\right)/2 +1}
    \right]
    \end{aligned}
\end{align*}
With numerical values, this is
\begin{align*}
    D_x(X = \Omega^E_{S,x}/\omega ) 
    =
    5 \Big[
    & -3~194~880 \sqrt{5} \exp{-6 \sqrt{5} \abs{X}}
    +83~187 \exp{-20 \abs{X}}
    \\
    &
    +16~046~875 \exp{-12 \abs{X}}
    -11~649~024 \exp{-10 \abs{X}}
    \\
    &
    + 110 \exp{- 8 \sqrt{10} \abs{X}} \Big(
    9~827 \sqrt{10}  \cosh(20 \abs{X})
    + 31~076 \sinh(20 \abs{X})
    \Big)
    \Big] \Big/ 2~892~032
\end{align*}
The magnitude distribution is finally
\begin{align}
\begin{split}
    D(X = \Omega^E_S/\omega) 
    =
    25 X \Big[
    & -4~792~320 \exp{-6 \sqrt{5} X}
    + 83~187 \exp{-20 X}
    \\
    &
    +9~628~125 \exp{-12 X}
    -5~824~512 \exp{-10 X}
    \\
    &
    + 905~520 \cosh(20X) \exp{-8\sqrt{10}X}
    \\
    &
    + 286~374 \sqrt{10} \sinh(20X) \exp{-8\sqrt{10}X}
    \Big]
    \Big/ 361~504
\end{split}
\end{align}

\subsection{Two-point spatial correlation tensors}

To investigate further the different behaviours of the three currents, we compute two-point space correlation functions for their components. For free monochromatic fields, cycle-averaging Poynting's theorem implies that $\div{\vb{P}} = 0$. Since we always have $\div{\vb{P_S}} = 0$, we also find $\div{\vb{P_O}} = 0$. In our system, these three vector fields are also homogeneous and isotropic, and therefore, their spatial correlation tensors are all of the form
\begin{align}
    \expval{u_i(\vb{0}) u_j(\vb{r})}
    =
    - \frac{1}{2} r f'(r) \frac{r_i r_j}{r^2} + \delta_{ij} \left[ f(r) + \frac{1}{2} r f'(r) \right]
\end{align}
and only the longitudinal correlation function $f(r)$ is required to describe the full tensor. This reasoning also applies to the electric and magnetic fields themselves, for which the longitudinal and lateral correlation functions are denoted $L(r)$ and $T(r)$ and are tabulated above. We will be able to express all correlation functions using these two elementary functions. \\

All the calculations in this section will involve taking averages of products of four Gaussian variables. We use the previously tabulated results and apply the following strategies:
\begin{itemize}
\item We note that the global phase shift $\vb{E} \rightarrow i \vb{E}$ amounts to $\vb{p^E} \rightarrow \vb{q^E}$ and  $\vb{q^E} \rightarrow -\vb{p^E}$ and should leave statistics unchanged (the same of course goes for the magnetic variables). Hence many averages are found to be identical by the appropriate exchange of $p$ and $q$ variables.

\item Most importantly, we use Isserlis' theorem for moments of Gaussian variables \cite{goodman_1985_statistical}
\begin{align}
    \expval{u_1u_2u_3u_4} = 
    \expval{u_1u_2}\expval{u_3u_4}
    + \expval{u_1u_3}\expval{u_2u_4}
    + \expval{u_1u_4}\expval{u_2u_3}
\label{eq:isserlis}
\end{align}
In our cases, the first term will involve local averages only, and will always vanish. The second term will contain correlations of the electric and magnetic fields separately, while the third term will contain correlations between the electric and magnetic variables. 
\end{itemize}

\subsubsection*{Poynting correlation tensor}

The longitudinal correlation function is
\begin{align*}
    f_P(r) & = \expval{P_x(\vb{0}) P_x(r\vb{e_x})}
    = 
    \sum_{j,k,l,m}
    \expval{ \frac{1}{4} \epsilon_{xjk} \epsilon_{xlm} 
    [p^E_j p^H_k + q^E_j q^H_k](\vb{0})
    [p^E_l p^H_m + q^E_l q^H_m](r\vb{e_x})}
\end{align*}
Using the aforementioned techniques, we can eventually express $f_P(r)$ using tabulated correlation functions
\begin{align*} 
    f_P(r)
    & = 
    \expval{  
    p^E_y(\vb{0}) p^E_y(r\vb{e}_x) 
    }^2
    +
    \expval{ 
    p^E_y(\vb{0}) q^H_z(r\vb{e}_x)
    }^2
    \\
    & =
    T^2(r) + \frac{(kr)^2}{4} L^2(r)
    \\
    & =
    \frac{2 k^4 r^4+\left(2 k^2 r^2-1\right) \cos (2 k r)-2 k r \sin (2 k r)+1}{8 k^6 r^6}
\end{align*}
The corresponding correlation tensor is
\begin{align}
\begin{split}
    \expval{P_i(\vb{0}) P_j(\vb{r})}
    & =
    \Bigg[
    \frac{2 R^4 +2 R \left(R^2-3\right) \sin (2 R)+\left(6 R^2-3\right) \cos (2 R)+3}{8 R^6} 
    \Bigg] \frac{r_i r_j}{r^2}
    \\
    & +
    \Bigg[
    \frac{R \left(2-R^2\right) \sin (2 R)+\left(1-2 R^2\right) \cos (2R)-1}{4 R^6}
    \Bigg] \delta_{ij}
\end{split}
\end{align}
with $R = kr$.

\subsubsection*{Orbital momentum correlation tensor}

The same strategy applies. We have
\begin{align*}
    f_O(r) & = \expval{ P^E_{O,x}(\vb{0})  P^E_{O,x}(r\vb{e}_x) } 
    = 
    \frac{1}{\omega^2}
    \sum_{j, k}
    \expval{ \frac{1}{4}
    [p^E_j \partial_x q^E_j - q^E_j \partial_x p^E_j](\vb{0})
    [p^E_k \partial_x q^E_k - q^E_k \partial_x p^E_k](r\vb{e}_x)
    }
\end{align*}
Considering the terms that are equal upon exchange of $p$ and $q$, we have
\begin{align*}
    \omega^2 f_O(r)
    & = 
    \sum_{j, k}
    \frac{1}{2}
    \expval{
    [p^E_j \partial_x q^E_j](\vb{0})
    [p^E_k \partial_x q^E_k](r\vb{e}_x)
    }
    - 
    \frac{1}{2}
    \expval{
    [p^E_j \partial_x q^E_j](\vb{0})
    [q^E_k \partial_x p^E_k](r\vb{e}_x)
    }
\end{align*}
We now use \eqref{eq:isserlis}. The first term involving local averages vanishes, and from now on we will drop the spatial dependence~: it is understood that in each average, the first variable is evaluated at $\vb{0}$ and the second at $r\vb{e}_x$.
\begin{align*}
    & = 
    \sum_{j, k}
    \frac{1}{2}
    \expval{
    p^E_j p^E_k
    }
    \expval{
    \partial_x q^E_j \partial_x q^E_k
    }
    +
    \frac{1}{2}
    \expval{
    p^E_j \partial_x q^E_k
    }
    \expval{
    \partial_x q^E_j p^E_k
    }
    - 
    \frac{1}{2}
    \expval{
    p^E_j q^E_k
    }
    \expval{
    \partial_x q^E_j \partial_x p^E_k
    }
    -
    \frac{1}{2}
    \expval{
    p^E_j \partial_x p^E_k
    }
    \expval{
    \partial_x q^E_j q^E_k
    }
\end{align*}
From tabulated results the third term is zero
\begin{align*}
    & = 
    \sum_{j, k}
    \frac{1}{2}
    \expval{
    p^E_j p^E_j
    }
    \expval{
    \partial_x q^E_j \partial_x q^E_j
    }
    +
    \frac{1}{2}
    \expval{
    p^E_j \partial_x q^E_k
    }
    \expval{
    \partial_x q^E_j p^E_k
    }
    -
    \frac{1}{2}
    \expval{
    p^E_j \partial_x p^E_k
    }
    \expval{
    \partial_x q^E_j q^E_k
    }
\end{align*}
Correlators involving derivatives are non-zero only if the corresponding correlator of components is non-zero. We have
\begin{align*}    
    \omega^2 f_O(r) & = 
    \sum_{j}
    -
    \frac{1}{2}
    \expval{
    p^E_j p^E_j
    }
    \partial^2_x
    \expval{
    p^E_j p^E_j
    }
    +
    \frac{1}{2}
    \left(
    \partial_x
    \expval{
    p^E_j p^E_j
    }
    \right)^2
    \\
    & = 
    \frac{1}{2}
    \left[
    (L')^2
    -
    LL''
    \right]
    +
    \left[
    (T')^2
    -
    TT''
    \right]
    \\
    & = k^2 \frac{2 k^6 r^6-3 k^4 r^4-4 k^2 r^2+2 k r \left(9-2 k^2 r^2\right) \sin (2 k r)+\left(k^4 r^4-14 k^2 r^2+9\right) \cos (2 k r)-9}{8 k^8 r^8}
\end{align*}
The correlation tensor is
\begin{align}
\begin{split}
    & \expval{P^E_{O,i}(\vb{0}) P^E_{O,j}(\vb{r})} =
    \\
    &
    \Bigg[ 
    \frac{ \left(R^6-3 R^4-6 R^2+\frac{1}{2} \left(R^4-24 R^2+72\right) R \sin (2 R)+3 \left(R^4-10 R^2+6\right) \cos (2 R)-18\right)}{4 R^8}
    \Bigg] \frac{r_i r_j}{r^2}
    \\
    & +
    \Bigg[ 
    \frac{\left(3 R^4+8 R^2-\left(R^4-20 R^2+54\right) R \sin (2 R)+\left(-5 R^4+46 R^2-27\right) \cos (2 R)+27\right)}{8 R^8}
    \Bigg] \delta_{ij}
\end{split}
\end{align}

\subsubsection*{Spin momentum correlation tensor}

The same strategy again applies. We have
\begin{align*}
    f_S(r) & = \expval{ P^E_{S,x}(\vb{0}) P^E_{S,x}(r \vb{e}_x) }
    = \frac{1}{\omega^2}
    \sum_{j, k}
    \frac{1}{4}
    \expval{
    [p^E_j \partial_j q^E_x - q^E_j \partial_j p^E_x](\vb{0})
    [p^E_k \partial_k q^E_x - q^E_k \partial_k p^E_x](r\vb{e}_x)
    }
\end{align*}
Noticing again that half of the averages are the same upon exchange of $p$ and $q$
\begin{align*}
    \omega^2 f_S(r)
    & = \sum_{j, k}
    \frac{1}{2}
    \expval{
    [p^E_j \partial_j q^E_x](\vb{0})
    [p^E_k \partial_k q^E_x](r\vb{e}_x)
    }
    -
    \frac{1}{2}
    \expval{
    [p^E_j \partial_j q^E_x](\vb{0})
    [q^E_k \partial_k p^E_x](r\vb{e}_x)
    }
\end{align*}
We use \eqref{eq:isserlis}, as before the term with local averages vanishes and we are left with
\begin{align*}
    \omega^2 f_S(r)
    & = 
    \sum_{j,k}
    \frac{1}{2}
    \expval{
    p^E_j p^E_k
    }
    \expval{
    \partial_j q^E_x \partial_k q^E_x
    }
    +
    \frac{1}{2}
    \expval{
    p^E_j \partial_k q^E_x
    }
    \expval{
    \partial_j q^E_x p^E_k
    }
    -
    \frac{1}{2}
    \expval{
    p^E_j q^E_k
    }
    \expval{
    \partial_j q^E_x \partial_k p^E_x
    }
    -
    \frac{1}{2}
    \expval{
    p^E_j \partial_k p^E_x
    }
    \expval{
    \partial_j q^E_x q^E_k
    }
\end{align*}
Using tabulated results some averages are zero
\begin{align*}
    \omega^2 f_S(r)
    & = 
    \sum_{j, k}
    \frac{1}{2}
    \expval{
    p^E_j p^E_j
    }
    \expval{
    \partial_j p^E_x \partial_j p^E_x
    }
    -
    \frac{1}{2}
    \expval{
    p^E_j \partial_k p^E_x
    }
    \expval{
    \partial_j p^E_x p^E_k
    }
\end{align*}
and we are left with evaluating the correlators involving spatial derivatives. They are tabulated in a section above, and we find
\begin{align*}
    \omega^2 f_S(r) 
    & = 
    \frac{1}{2}
    \expval{
    p^E_x p^E_x
    }
    \expval{
    \partial_x p^E_x \partial_x p^E_x
    }
    +
    \expval{
    p^E_y p^E_y
    }
    \expval{
    \partial_y p^E_x \partial_y p^E_x
    }
    +
    \frac{1}{2}
    \expval{
    p^E_x \partial_x p^E_x
    }^2
    +
    \expval{
    p^E_y \partial_y p^E_x
    }^2
    \\
    & = 
    - \frac{1}{2}
    L(r)
    L''(r)
    -
    T(r)
    \frac{2 L'(r)}{r}
    +
    \frac{1}{2}
    {L'}^2(r)
    +
    \frac{1}{4}
    {L'}^2(r)
    \\
    & = 
    \frac{3}{4}
    {L'}^2(r)
    - \frac{L(r) L''(r)}{2}
    -
    2 L'(r)
    \frac{T(r)}{r}
    \\
    & =
    k^2 \frac{k^4 r^4-k^2 r^2+2 k r \left(9-5 k^2 r^2\right) \sin (2 k r)+\left(3 k^4 r^4-17 k^2 r^2+9\right) \cos (2 k r)-9}{8 k^8 r^8}
\end{align*}
and the correlation tensor is
\begin{align}
\begin{split}
    & \expval{P^E_{S,i}(\vb{0}) P^E_{S,j}(\vb{r})} \\
    & =
    \frac{r_i r_j}{r^2}
    \Bigg[
    \frac{\left(2 R^4-3 R^2+ 3\left(R^4-14 R^2+24\right) R \sin (2 R)+\left(16 R^4-69 R^2+36\right) \cos (2 R)-36\right)}{8 R^8}
    \Bigg]
    \\
    & +
    \delta_{ij}
    \Bigg[
    \frac{ \left(-R^4+2 R^2+\left(-3 R^4+32 R^2-54\right) R \sin (2 R)+\left(-13 R^4+52 R^2-27\right) \cos (2 R)+27\right)}{8 R^8}
    \Bigg]
\end{split}
\end{align}

\subsubsection*{Scalar current correlation tensor}

The complex scalar random field $\Psi$ can be defined as a sum of scalar waves as follows
\begin{align}
    \Psi = \sqrt{\frac{2}{N}} \sum_{n=1}^N e^{i \vb{k}_n \vdot \vb{r} + i \psi_n}
    =
    p^\Psi + i q^\Psi
\end{align}
$\Psi$ obeys Helmholtz's equation with wavevector $k$. The current in this case is 
\begin{align}
    \vb{J} = \frac{1}{2\omega} \Im{ \Psi^* \grad \Psi}
\end{align}
and it is divergenceless \cite{berry_2009_optical}, such that the previous strategy again applies. We have
\begin{align*}
    f_J(r) = \expval{J_x(\vb{0}) J_x(r\vb{e}_x)}
    =
    \frac{1}{\omega^2}
    \expval{ \frac{1}{4} [p^\Psi\partial_x q^\Psi - q^\Psi\partial_x p^\Psi](\vb{0}) 
    [p^\Psi\partial_x q^\Psi - q^\Psi\partial_x p^\Psi](r\vb{e}_x)}
\end{align*}
using the same procedure, we find
\begin{align*}
    \omega^2 f_J(r) = \frac{1}{2}
    \left[{C'}^2(r)-C(r)C''(r)\right]
    =
    \frac{ \left(2 R^2+\cos (2 R)-1\right)}{4 R^4}
\end{align*}
producing the correlation tensor
\begin{align}
\begin{split}
    & \expval{J_{i}(\vb{0}) J_{j}(\vb{r})} 
    =
    \frac{r_i r_j}{r^2}
    \Bigg[
    \frac{\left(2 R^2+R \sin (2 R)+2 \cos (2 R)-2\right)}{4 R^4}
    \Bigg]
    +
    \delta_{ij}
    \Bigg[
    -\frac{ (R \sin (2 R)+\cos (2 R)-1)}{4 R^4}
    \Bigg]
\end{split}
\end{align}

\bibliography{References}